\documentclass[]{aa}

\usepackage{graphicx}
\usepackage{txfonts}
\usepackage{mathtools}
\usepackage{amsmath}
\usepackage{pdfpages}
\usepackage{natbib}
\usepackage{url}
\usepackage{placeins}
\usepackage{tikz}
\usepackage{wasysym}
\usepackage{gensymb}
\usetikzlibrary{shapes,arrows}

\newcommand{\eq}{Eq.}

\newcommand{\fig}{Figure}

\newcommand{\Bv}{\textbf{B}}
\newcommand{\vv}{\textbf{v}}
\newcommand{\rv}{\textbf{r}}

\begin{document}

   \title{Smoothed particle magnetohydrodynamics with the geometric density average force expression}

   \author{Robert Wissing \and Sijing Shen
          }
   \institute{
   Institute of Theoretical Astrophysics, University of Oslo, Postboks 1029, 0315 Oslo, Norway \\ \email{robertwi@astro.uio.no; sijing.shen@astro.uio.no;}}
   \date{}
   \titlerunning{SPMHD with the geometric density average force expression}
  \authorrunning{Wissing \& Shen}
 
  \abstract
   { 
We present a novel method of magnetohydrodynamics (MHD) within the smoothed particle hydrodynamics scheme (SPMHD) using the geometric density average force expression. Geometric density average within smoothed particle hydrodynamics (GDSPH) has recently been shown to reduce the leading order errors and greatly improve the accuracy near density discontinuities, eliminating surface tension effects. Here, we extend the study to investigate how SPMHD benefits from this method. We implement ideal MHD in the {\sc Gasoline2} and {\sc Changa} codes with both GDSPH and traditional smoothed particle hydrodynamics (TSPH) schemes. A constrained hyperbolic divergence cleaning scheme was employed to control the divergence error and a switch for artificial resistivity with minimized dissipation was also used. We tested the codes with a large suite of MHD tests and showed that in all problems, the results are comparable or improved over previous SPMHD implementations. While both GDSPH and TSPH perform well with relatively smooth or highly supersonic flows, GDSPH shows significant improvements in the presence of strong discontinuities and large dynamic scales. In particular, when applied to the astrophysical problem of the collapse of a magnetized cloud, GDSPH realistically captures the development of a magnetic tower and jet launching in the weak-field regime, while exhibiting fast convergence with resolution, whereas TSPH failed to do so. Our new method shows qualitatively similar results to those of the meshless finite mass/volume (MFM/MFV) schemes within the {\sc Gizmo} code, while remaining computationally less expensive.
  }


   \maketitle
%

\section{Introduction}
Magnetic fields are important in a wide array of different astrophysical systems. In star formation, they govern the dynamics at several stages during collapse.  They are critical in the launching of jets from a broad range of sources.  They also play a major role in the transport of angular momentum in ionized accretion disks due to the magnetorotational instability. Magnetic fields have been largely neglected in galaxy formation simulations, mostly due to the technical difficulties associated with them. It is only recently that researchers have begun to apply them \citep{2009ApJ...696...96W,2011MNRAS.415.3189K,2013MNRAS.432..176P,2016MNRAS.457.1722R,2017ApJ...843..113B,2017MNRAS.471..144S,2017MNRAS.469.3185P,2019MNRAS.483.1008S}. The importance of magnetic fields in galaxy formation is clear from observations of the Milky Way and nearby galaxies, which reveal that the magnetic energy is in equipartition with the thermal and turbulent energies \citep{1990ApJ...365..544B,1996ARA&A..34..155B}. This means that they are likely to have a large dynamical effect on the evolution of the galaxy, adding significant non-thermal pressure that can suppress star-formation \citep{2013MNRAS.432..176P}. In addition, it has been shown that magnetic fields have a strong impact on fluid instabilities \citep{1995ApJ...453..332J,2015MNRAS.449....2M}, which may affect how gas in the intergalactic medium (IGM) accretes onto galaxies and how gas in galactic outflows leaves (or cycles back to) galaxies. The strength and structure of magnetic fields in galaxies also determine the transport of cosmic rays (CRs), which has recently emerged as a promising candidate for driving galactic outflows because they have long cooling time scales \citep{2012MNRAS.423.2374U,2013ApJ...777L..16B,2016ApJ...824L..30P,2018ApJ...868..108B}.
\\ \\ \\
Apart from the improvements in the general method, advances have been made in the magnetohydrodynamics extension for smoothed particle hydrodynamics (SPMHD). The modern foundation of SPMHD comes largely from the work of \cite{2004MNRAS.348..139P} which was built on the earlier work of \cite{1985MNRAS.216..883P}. The two main technical difficulties to overcome for SPMHD, are the handling of divergence errors and the choice of an artificial resistivity term to capture shocks and discontinuities in the magnetic field.
\\ \\
Artificial dissipation terms are required to smooth out discontinuities in any fluid quantity in all numerical hydrodynamics methods. In SPH, this is most commonly achieved via explicit artificial dissipation. To avoid excessive dissipation away from shocks and discontinuities, switches have been developed to limit where the artificial dissipation terms are active. For magnetic fields, newly developed artificial resistivity switches \citep{2018PASA...35...31P,2013MNRAS.436.2810T} have significantly reduced the amount of dissipation and improved the method in the weak field regime. 
\\ \\
Unphysical divergence errors (magnetic monopoles) can arise from the discretization and numerical integration of the MHD equations. Divergence errors in SPMHD for magnetic-dominated scenarios need to be handled with care, as they can produce a negative force between particles which leads to the tensile instability \citep{2000JCoPh.159..290M}. As such, the force produced from the divergence needs to be partly removed in the strong field regime for the method to remain stable, this breaks momentum and energy conservation in proportion to the divergence error. It is therefore crucial to try to keep the divergence error as close to zero as possible. While grid codes have access to the constrained transport scheme \citep{1988ApJ...332..659E}, which ensures a divergence free field up to machine precision, it cannot easily be implemented within meshless methods, due to the absence of regular spatial grid surfaces. Generation of divergence free fields in SPMHD have been explored in detail, however, all of them suffer from problems. Generation of magnetic fields from Euler potentials ($B = \nabla \alpha \times \nabla \beta $) cannot wind the magnetic field and thereby not produce a dynamo \citep{2010MNRAS.401..347B}. \cite{2010MNRAS.401.1475P} showed that vector potential implementations ($B = \nabla \times A$) are plagued with numerical instabilities. However,  \cite{2015JCoPh.282..148S} recently showed that with additional diffusion, smoothing of the magnetic field, and enforcing the Coulomb gauge ($\nabla\cdot A = 0$), the vector potential formalism could remain stable for a handful of test cases. Additional testing would be required to determine the robustness of the method. The most popular method to deal with divergence error in meshless methods, is to evolve the magnetic field via the induction equation and then to "clean" the divergence away. In general, this is done by introducing a separate scalar field which couples to the induction equation such that it produces a damped wave equation for the divergence error, so the divergence is spread outward like a damped wave. The method was first developed in \cite{2002JCoPh.175..645D} and was improved by \cite{2012JCoPh.231.7214T}, who introduced a constrained version of the method.  This ascertains that the magnetic energy is either conserved or dissipated. This was updated in \cite{2016JCoPh.322..326T} to correctly allow variable cleaning speed, which further improved the method.
\\ \\
These new improvements in artificial dissipation and divergence error controlling have significantly increased the accuracy and convergence of the SPMHD method. There have also been implementations of non-ideal MHD in SPMHD proposed recently \citep{2013MNRAS.434.2593T,2014MNRAS.444.1104W,2015ApJ...810L..26T,2015MNRAS.452..278T,2016MNRAS.457.1037W,2018PASA...35...31P}, which include Ohmic resistivity, ambipolar diffusion, and the Hall effect. 
\\ \\
As mentioned previously, the numerical surface tension seen in traditional SPH (TSPH) can be solved by using a different gradient operator (GDSPH) \citep{2017MNRAS.471.2357W}. This substantially improves the accuracy of pressure forces across density jumps and provides a more physical form for the internal energy equation, where it represents a direct discretization of $\frac{du}{dt}=-\frac{P}{\rho} \ \nabla \cdot v $ from the Euler equations while retaining all the usual conservation properties. In \citet{2017MNRAS.471.2357W}, the authors show that GDSPH, together with an explicit turbulent diffusion term on thermal energy, yields excellent results in fluid mixing test cases, such as the Kelvin-Helmholtz instability and the blob test.    
\\ \\
In this paper, we investigate how SPMHD benefits from the use of GDSPH. As such, we have implemented MHD within the {\sc Gasoline2} \citep{2017MNRAS.471.2357W} and {\sc Changa} \citep{2015ComAC...2....1M} codes, which both utilize the GDSPH formalism. {\sc Gasoline2} is a highly parallel, state-of-the-art code for cosmological structure formation simulations which includes all the features of modern SPH methods. {\sc Changa} includes all the same SPH methods as {\sc Gasoline2}, but it is written in an inherently parallel language {\sc Charm++} \citep{45656fcdf4f648b48a97b0ecec6177c0} which enables more efficient parallelization. The major difference between the two codes lies in the gravity solver, which is different in {\sc Changa} because it uses an oct-tree, rather than an arbitrary binary KD-tree as in {\sc Gasoline2}.
\\ \\
This paper is organized as follows. In Section~\ref{sec:theory}, we go through the SPMHD theory and show how the equations can be formulated using the GDSPH approach. In Section~\ref{sec:tests}, we test our implementation on a large suite of standard test cases and in Section~\ref{sec:collapse} we apply the code to an astrophysical application: the collapse of a magnetized cloud. In Section~\ref{sec:discussion}, we discuss our results and present some concluding remarks.
\section{Theory}
\label{sec:theory}
In this section, we show how the MHD equations can be formulated in a conservative way within the GDSPH framework.  The development is similar to the findings in previous work \citep{2004MNRAS.348..139P,2012JCoPh.231..759P,2016JCoPh.322..326T,2018PASA...35...31P}  and we direct the reader to these papers for additional background details.
\subsection{MHD theory}
\label{subsec:mhdtheory}
The two main equations which are relevant for ideal MHD are the Lorentz force law and the induction equation. Assuming that the fluid is an ideal conductor ($E=0$), the Lorentz force law can be written as:
\begin{equation}
\label{eq:momlorentz}
\frac{d\vv}{dt}=\frac{1}{\mu_0 \rho}(\nabla \times \Bv)\times \Bv=\frac{1}{\mu_0 \rho}\left(-\frac{1}{2}\nabla \Bv^2+(\Bv\cdot\nabla)\Bv\right) ,
\end{equation}
where $\vv$, $\rho$, $\Bv$ and $\mu_0$ is the velocity, density, magnetic field and vacuum permeability, respectively. The first term acts like an isotropic magnetic pressure term, while the other term acts as an attractive term along magnetic field lines (tension). Going forward, we define code units such that $\mu_0=1$. The conservative form of SPMHD is attained by using the stress tensor to describe the momentum equation. Assuming that the magnetic field is divergence free, the MHD stress tensor can be written as:
\begin{equation}
\label{eq:stress}
S^{ij}=-\delta^{ij}\left(P+\frac{\Bv^2}{2}\right)+B^iB^j ,
\end{equation}
where $P$ is the thermal pressure and $\delta^{ij}$ is the Kronecker delta. The momentum equation can then be written as:
\begin{equation}
\label{eq:momstress}
\frac{d\vv}{dt}=\frac{\nabla\cdot \textbf{S}}{\rho}=-\frac{1}{\rho}\nabla\left(P+\frac{\Bv^2}{2}\right)+\frac{1}{\rho}\left[(\Bv\cdot\nabla)\Bv+\Bv(\nabla\cdot \Bv)\right] .
\end{equation}
There is an extra tension force term which would normally have no effect due to the assumption $\nabla\cdot \Bv=0$. However, as mentioned in the introduction, this constraint is usually not fully upheld in numerical codes. To avoid numerical instability within SPH, this term needs to be negated when the magnetic pressure exceeds the thermal pressure. 

The change in the magnetic field is obtained from the induction equation:
\begin{equation}
\label{eq:induccont}
    \frac{d\Bv}{dt}=\nabla\times(\vv\times \Bv)=(\Bv\cdot\nabla)\vv-\Bv(\nabla\cdot \vv) ,
\end{equation}
where the first term affects the magnetic field through shearing motion, while the second will increase the magnetic field when undergoing compression. A combined effect of the two terms is to enhance the field due to compression perpendicular to the field direction (for example, $B \propto \rho^{2/3}$ for spherical collapse). Compression in the direction of the field has no effect.

\subsection{SPH discretization}
\label{subsec:sphdisc}
Derivatives within SPH can be discretized in a number of ways, and a general formulation is given by \cite{2012JCoPh.231..759P}:
 \begin{equation}
 \label{eq:symoperatorgen}
     \frac{\nabla A}{\rho} = \frac{\phi}{\rho}\left[\frac{A}{\phi^2}\nabla \phi + \nabla\left(\frac{ A}{\phi}\right)\right] \approx \sum_b \frac{m_b}{\rho_a\rho_b}\left(A_a\frac{\phi_b}{\phi_a}+A_b\frac{\phi_a}{\phi_b}\right)\nabla_a \overline{W}_{ab} ,
\end{equation}
 \begin{equation}
  \label{eq:antisymoperatorgen}
     \frac{\nabla A}{\rho} = \frac{1}{\phi\rho} \left[\nabla (\phi A) - A\nabla \phi\right] \approx \sum_b \frac{m_b}{\rho_a\rho_b}\frac{\phi_b}{\phi_a}\left(A_b-A_a\right)\nabla_a \overline{W}_{ab},
\end{equation}
where $\phi$ can be any arbitrary, differentiable scalar quantity. The geometric density average force formulation (GDSPH) corresponds to using $\phi=1$ while traditional SPH corresponds to using $\phi=\rho$. GDSPH therefore gives the following symmetric and anti-symmetric gradient operators:
 \begin{equation}
 \label{eq:symoperator}
     \frac{\nabla A}{\rho} \approx \frac{1}{\rho}(A\nabla 1 + \nabla A) = \sum_b \frac{m_b}{\rho_a\rho_b}\left(A_a+A_b\right)\nabla_a \overline{W}_{ab} ,
\end{equation}
 \begin{equation}
  \label{eq:antisymoperator}
     \frac{\nabla A}{\rho} \approx \frac{1}{\rho} (\nabla A - A\nabla 1) = \sum_b \frac{m_b}{\rho_a\rho_b}\left(A_b-A_a\right)\nabla_a \overline{W}_{ab} .
\end{equation}
Here, $\nabla_a\overline{W}_{ab}$ is a symmetric gradient of the smoothing kernel:
\begin{equation}
\label{eq:smkern}
\nabla_a\overline{W}_{ab}=\frac{1}{2}\left[f_a\nabla_a W(r_{ab},h_a)+f_b\nabla_b W(r_{ab},h_b)\right] ,
\end{equation}
where $W$ is the smoothing kernel, $h_a$ is the smoothing length of particle $a$, and $r_{ab}=|\mathbf{r_a}-\mathbf{r_b}|$ is the distance between particle $a$ and $b$. Here, $f_a$ is a correction term introduced in \cite{2017MNRAS.471.2357W} to ensure that internal energy and density evolve consistently, such that entropy is tightly conserved. To attain a conservative formalism for SPH, the symmetric gradient operator is applied to the equations of motion and the anti-symmetric gradient operator is applied to the internal energy equation \footnote{This can clearly be seen when deriving the SPH equations from the least action principle \citep{2012JCoPh.231..759P}.}. As a consequence, zeroth order errors arise in the equations of motions which will depend on the local particle distribution \footnote{This can be seen as an inherent re-meshing procedure, where the particles try to arrange themselves to maximize the sum of the particle volumes and reach a minimum energy state.}. A generalized error term for the zeroth order errors is given by:
\begin{equation}
    \label{eq:e0error}
     \textbf{E}_0= \sum_b \frac{m_b}{\rho_b}\left(\Phi_{ab}+\Phi_{ab}^{-1}\right)\nabla_a \overline{W}_{ab} ,
\end{equation}
where $\Phi=\frac{\phi_a}{\phi_b}$ depend on the chosen scalar quantity $\phi$ in \eq~\ref{eq:symoperatorgen} and \ref{eq:antisymoperatorgen}. As shown by \cite{2010MNRAS.405.1513R}, in TSPH $\Phi_{ab}=\frac{\rho_a}{\rho_b}$, while in GDSPH $\Phi_{ab}=1$. It is then evident that these errors are more severe at density gradients in TSPH than in GDSPH (where they are explicitly independent of the density gradient). A similar improvement can be seen for the linear errors.
\\ \\
Applying the symmetric gradient operator to the momentum equation (\eq \ref{eq:momstress}) and the anti-symmetric gradient operator to the induction equation (\eq\ref{eq:induccont}) gives:
\begin{equation}
  \label{eq:momdisc}
    \frac{d v_a^i}{dt}=\sum_b \frac{m_b}{\rho_a\rho_b}\left(S_a^{ij}+S_b^{ij}\right)\nabla_a^j \overline{W}_{ab} + f_{divB,a}^i \ ,
\end{equation}
\begin{equation}
  \label{eq:inducdisc}
    \frac{d\Bv_a}{dt}=\sum_b \frac{m_b}{\rho_b}\left[\Bv_a(\vv_{ab}\cdot\nabla_a \overline{W}_{ab})-\vv_{ab}(\Bv_a\cdot\nabla_a \overline{W}_{ab})\right] ,
\end{equation}
where $\vv_{ab}=\vv_a-\vv_b$. The stability term $f_{divB,a}^i$ is added to avoid the tensile instability. This can occur due to divergence errors when the magnetic pressure exceeds the gas pressure ($\frac{B^2}{2}>P$) \citep{1985MNRAS.216..883P}. The stability term is defined as: 
\begin{equation}
\label{eq:stabilityterm}
    f_{divB,a}^i=-\hat{B_a^i}\sum_b \frac{m_b}{\rho_a\rho_b}\left(\textbf{B}_a+\textbf{B}_b\right)\cdot\nabla_a \overline{W}_{ab} .
\end{equation}
This basically removes the divergence term ($-\frac{\Bv}{\rho}\nabla\cdot \Bv$) from \eq \ref{eq:momstress} \citep{2001ApJ...561...82B,2012JCoPh.231..759P}.  Removing a term from the conservative momentum equation effectively breaks momentum conservation.  However, the error introduced will be proportional to the divergence. To minimize its effect in the weak field regime, we use the scheme from \cite{2004ApJS..153..447B} with a factor of $\hat{B_a^i}=B_a^i$ for $\beta<1$ as advocated by \cite{2012JCoPh.231.7214T}:
\begin{equation}
\label{eq:borvescheme}
\hat{B_a^i}=
\begin{cases}
B_a^i \quad  & \beta<1\\
B_a^i(2-\beta) & 1<\beta<2\\
0 & \mbox{otherwise} ,
\end{cases}
\end{equation}
where $\beta=\frac{2P}{B^2}$ is the plasma beta . 
\subsection{Treating discontinuities}
When fluid quantities become discontinuous, they are no longer differentiable, which is problematic as differentiability is assumed by the SPMHD equations. Artificial resistivity is required to smooth out discontinuities in the magnetic field which can occur both along and orthogonal to the fluid motion and in both compression and rarefaction. The artificial resistivity can be represented as an isotropic diffusion: 
\begin{equation}
    \frac{d\Bv}{dt}_{diss}=\eta\nabla^2\Bv ,
\end{equation}
where $\eta$ is a resistivity parameter.
We use the Brookshaw method \citep{1985PASAu...6..207B}, which estimates the second derivative by using the first derivative kernel and the difference in the field divided by the particle spacing. Following the GDSPH discretization, we get: 
\begin{equation}
    \frac{d\Bv_a}{dt}_{diss}=\sum_b \frac{m_b}{\rho_b}\left(\frac{\eta_a+\eta_b}{|{\rv}_{ab}|}\right)\Bv_{ab}\left(\hat{\rv}_{ab}\cdot\nabla_a \overline{W}_{ab}\right) , 
\end{equation}
where ${\Bv}_{ab}={\Bv}_a-{\Bv}_b$ and $\hat{\rv}_{ab}={\rv}_{ab}/|{\rv}_{ab}|$. To conserve energy, the change in the internal energy becomes:
\begin{equation}
    \frac{du_a}{dt}_{diss}=-\frac{1}{2}\sum_b \frac{m_b}{\rho_a\rho_b}\left(\frac{\eta_a+\eta_b}{|{\rv}_{ab}|}\right)\Bv_{ab}^2\left(\hat{\rv}_{ab}\cdot\nabla_a \overline{W}_{ab}\right) .
\end{equation}
To reduce dissipation away from shocks, we can introduce a varying and resolution dependent resistivity parameter:
\begin{equation}
\label{eq:eta}
\eta=\frac{1}{2}\alpha_B v_{sig,B} |{\rv}_{ab}| ,
\end{equation}
where $\alpha_B$ is a dimensionless coefficient and $v_{sig,B}$ is the signal speed. Proper choice of $\alpha_B$ and $v_{sig,B}$ makes the artificial resistivity second order accurate away from shocks ($\eta \propto h^2$). We choose to implement the resistivity from {\sc Phantom} \citep{2018PASA...35...31P} where the signal speed is activated following: 
\begin{equation}
v_{sig,B}=|\vv_{ab}\times \hat{\rv}_{ab}| .
\label{eq:vsig}
\end{equation}
The dimensionless coefficient $\alpha_B$ is set to a constant. In {\sc Phantom}, this coefficient is set to $\alpha_B=1$, however, from our tests we find that $\alpha_B=0.5$ provides sufficient dissipation. This switch was shown to be the least dissipative compared to previous switches, while still capturing the correct magnetic features  \citep{2017arXiv170607721W}.

\subsection{Divergence cleaning}
\label{subsec:divclean}
As we discussed in the introduction, divergence errors are generated by the discretization and integration of the MHD equations. Apart from creating an unphysical magnetic field, it also forces us to introduce a stability term (\eq\ref{eq:stabilityterm}), which breaks momentum conservation in the strong field regime. This makes it crucial to reduce the divergence errors as much as possible. The best way found in SPMHD is by introducing a divergence cleaning scheme \citep{2012JCoPh.231.7214T}. In general, this is done by introducing a separate scalar field which couples to the induction equation, such that it produces a damped wave equation for the divergence error. That is, the divergence is spread outward like a damped wave. In our implementation, we employ the constrained hyperbolic divergence cleaning from \cite{2016JCoPh.322..326T}, an improved version of the method presented by \cite{2002JCoPh.175..645D}. The constrained hyperbolic divergence cleaning ensures that magnetic energy is either conserved or dissipated. In this method, a scalar field $\psi$ is coupled to the induction equation as follows:
\begin{equation}
\label{eq:Bdivclean}
    \left(\frac{d\Bv}{dt}\right)_{\psi}=-\nabla\psi.
\end{equation}
The scalar field $\psi$ evolves according to:
 \begin{equation}
\label{eq:BClean}
     \frac{d}{dt}\left(\frac{\psi}{c_h}\right)=-c_h\nabla\cdot \Bv-\frac{1}{\tau}\frac{\psi}{c_h}-\frac{1}{2}\psi (\nabla\cdot \vv).
 \end{equation}
where $\tau$ is the decay time and $c_h$ is the wave cleaning speed:
\begin{equation}
c_h=f_{clean}v_{mhd} ,
\end{equation}
\begin{equation}
\label{eq:vmhd}
v_{mhd}=\sqrt{c_s^2+v_A^2} ,
\end{equation}
\begin{equation}
v_A=\sqrt{\frac{B^2}{\rho}} .
\end{equation}
Here, $c_s$ is the speed of sound, $v_A$ the Alfv\'{e}n velocity, and $f_{clean}$ is an overcleaning factor. The $f_{clean}$ factor can be used to increase the amount of divergence cleaning, however, this will reduce the timestep\footnote{This is a significant increase in computational cost, so it is in general not recommended to use an $f_{clean}$ value above 1. But it does allows for a simple way to reduce the divergence error, if that is required.} according to $\Delta t \rightarrow \Delta t/f_{clean}$. Combining the cleaning equation with the induction equation produces a damped wave equation for the divergence (this form assumes constant $c_h$ and $\tau$):
 \begin{equation}
     \frac{\partial^2(\nabla\cdot \Bv)}{\partial t^2} -c_h^2\nabla^2(\nabla\cdot \Bv)+\frac{1}{\tau}\frac{\partial(\nabla\cdot \Bv)}{\partial t}=0 ,
 \end{equation}
 which effectively shows that the divergence is spread out and damped. The decay time is given by:
 \begin{equation}
     \tau_a=\frac{h_a}{c_{h,a}\sigma_c} .
 \end{equation}
 Here, $\sigma_c$ is a dimensional constant, and was shown to be optimal with a value of 1 in 3D. Following \cite{2012JCoPh.231.7214T}, $\nabla \psi$ is discretized using the symmetric gradient operator (\eq \ref{eq:symoperator}) and $\nabla \cdot B$ using the anti-symmetric gradient operator (\eq \ref{eq:antisymoperator}). Within the GDSPH discretization, \eq \ref{eq:Bdivclean} and \eq \ref{eq:BClean} become:
 \begin{equation}
     \left(\frac{d\Bv}{dt}\right)_{\psi,a}=-\sum_b \frac{m_b}{\rho_b}\left(\psi_a+\psi_b\right)\nabla_a \overline{W}_{ab} ,
 \end{equation}
 \begin{equation}
     \frac{d}{dt}\left(\frac{\psi}{c_h}\right)_a=c_h^a\sum_b \frac{m_b}{\rho_b}\Bv_{ab}\cdot\nabla_a \overline{W}_{ab}+\frac{\psi_a}{2c_h^a}\sum_b \frac{m_b}{\rho_b}\vv_{ab}\cdot\nabla_a \overline{W}_{ab}-\frac{\psi_a}{c_h^a\tau_a} .
 \end{equation}
The divergence cleaning dissipates energy from the magnetic field. However, this term is so small compared to the other dissipation terms that it is not worth accounting for. We could, of course, add this energy to heat and conserve energy, however, as discussed by \cite{2012JCoPh.231.7214T}, the removal of magnetic energy and  subsequent  generation  of  thermal  energy  would  be  non-local  due  to  the  coupling  of  parabolic  diffusion  with hyperbolic transport. Due to this, we simply removed the energy.
\\ \\
To ensure that simulations are not affected by the divergence error, we monitor the normalized divergence error:
\begin{equation}
\label{eq:divBerr}
\epsilon_{divB}=\frac{h|\nabla\cdot \Bv|}{|B|} .
\end{equation}
The mean of this quantity should preferably remain below $10^{-2}$. However, regions of locally high divergence error can occur, so careful inspection of the divergence error is required to ensure the quality of simulations.
\subsection{Shock capturing}
\label{subsec:shockcap}
To correctly capture shocks in MHD, we need to modify the artificial viscosity term in the momentum equation (see \cite{2017MNRAS.471.2357W} for a detailed description of the artificial viscosity term in {\sc Gasoline2}). For MHD the sound speed is replaced by the fast magnetosonic speed (\eq \ref{eq:vmhd}). We also modify the gradient-based shock detector introduced in {\sc Gasoline2}, which determines the direction of the shock from the pressure gradient. For the MHD, we must include the Lorentz force to correctly determine the direction of the shock. A more general way to determine the direction of the shock is to estimate the acceleration of the MHD forces without the dissipation terms before the actual force calculation:
\begin{equation}
\hat{\textbf{n}}=-\left(\frac{\frac{d\vv}{dt}}{\left|\frac{dv}{dt}\right|} \right)_{no \ diss} .
\end{equation}
This addition improves the behavior of shock detection in convergent flows for MHD. In {\sc Gasoline2}, the diffusion of fluid scalar variables such as thermal energy, metals and so forth are modeled using subgrid turbulent mixing \citep{2008MNRAS.387..427W,2010MNRAS.407.1581S}. However, we found that in strong shocks like the MHD blastwave, the thermal dissipation is not enough and can lead to incorrect velocity profiles. As such, we add a thermal shock dissipation similar to \eq 4.5 with \eq 4.8 in \cite{1992ARA&A..30..543M}, however, we use $\overline{c}_{ab}=0$ and with a larger constant (fitted parameter from the {\sc Gasoline2} code):
\begin{equation}
\frac{du}{dt}_{shock}=-\sum_b \frac{m_b}{\overline{\rho}_{ab}} d_{shock} u_{ab} \frac{\left(\hat{\textbf{r}}_{ab}\cdot\nabla_a \overline{W}_{ab}\right)}{\left|r_{ab}\right|} ,
\end{equation}
\begin{equation}
d_{shock}=16\overline{h}_{ab}\,|\mu_{ab}| ,
\end{equation}
\begin{equation}
    \mu_{ab}=
\begin{cases}
   \frac{\overline{h}_{ab}(\textbf{v}_{ab}\cdot \textbf{r}_{ab})}{r_{ab}^2+0.01\overline{h}_{ab}^2}  & \text{for}\ \textbf{v}_{ab}\cdot \textbf{r}_{ab} < 0 , \\
  0  & \text{otherwise,}
\end{cases}
\end{equation}
\section{Test problems}
\label{sec:tests}
In this section, we present the results from our test cases. All the simulations were run with the MHD version of {\sc Gasoline2}. To remain consistent and show the production quality of the method, we decided to run all the tests in 3D and with a default set of code parameters (described below). While glass-like initial conditions should always be used to correctly capture the natural state of 3D SPH simulations, for the sake of comparison, we elected to follow the initial setups from other authors, which often use lattice-based initial conditions. Test cases that are originally 1D or 2D are made 3D by extending the non-active dimensions by a set number of particles. By default, {\sc Gasoline2} sets the smoothing length based on a fixed number of neighbours. However, we found that in very uniform and precise tests, as in the circularized Alfv\'{e}n wave test, this approach generates small force errors that generate perturbations in the traveling wave. As such, we made the smoothing length directly proportional to the density and simultaneously determined the density and smoothing length using an iterative summation \citep{2002MNRAS.333..649S}. We note, however, that in all other tests, no visible effect was seen. We ran simulations with both TSPH and GDSPH, and the only difference between them is the choice of $\phi$ in \eq~\ref{eq:symoperatorgen} and \eq~\ref{eq:antisymoperatorgen}; all the other numerical schemes and parameters remain the same. In many of these tests, we compare the results to the state-of-the-art SPMHD code {\sc Phantom} \citep{2018PASA...35...31P}, and to the PSPH and the new meshless finite mass/volume (MFM/MFV) method of the {\sc Gizmo} code \citep{2015MNRAS.450...53H}. The MFM/MFV method utilizes a Lagrangian Godunov type method that employs more complex gradient operators and calculates fluxes from Riemann solvers.
\\ \\
\textbf{Default set of code parameters:}
\\
For the smoothing kernel, we used a Wendland C4 kernel \citep{Wendland1995} with 200 neighbours\footnote{Choice of kernel and neighbour number discussed at the end of section 3.1}. Our artificial viscosity (AV) followed the prescription given in \cite{2017MNRAS.471.2357W}, the AV parameters were set to $\alpha_{max}=4$, $\alpha_{min}=0$, $\tau=0.1 \frac{h}{c}$ and $\beta=2$. The artificial resistivity (AR) followed from the method outlined in Section~\ref{subsec:sphdisc}, and the AR parameter was set to $\alpha_B=0.5$. The thermal diffusion followed the turbulent mixing model described in \citet{2008MNRAS.387..427W} and \citet{2010MNRAS.407.1581S}, with the turbulent diffusion coefficient set to $C=0.03$.

\subsection{Circularized polarized Alfv\'{e}n wave}
\label{subsec:alfven}
\begin{figure}[!h]
    \centering
    \includegraphics[width=\hsize]{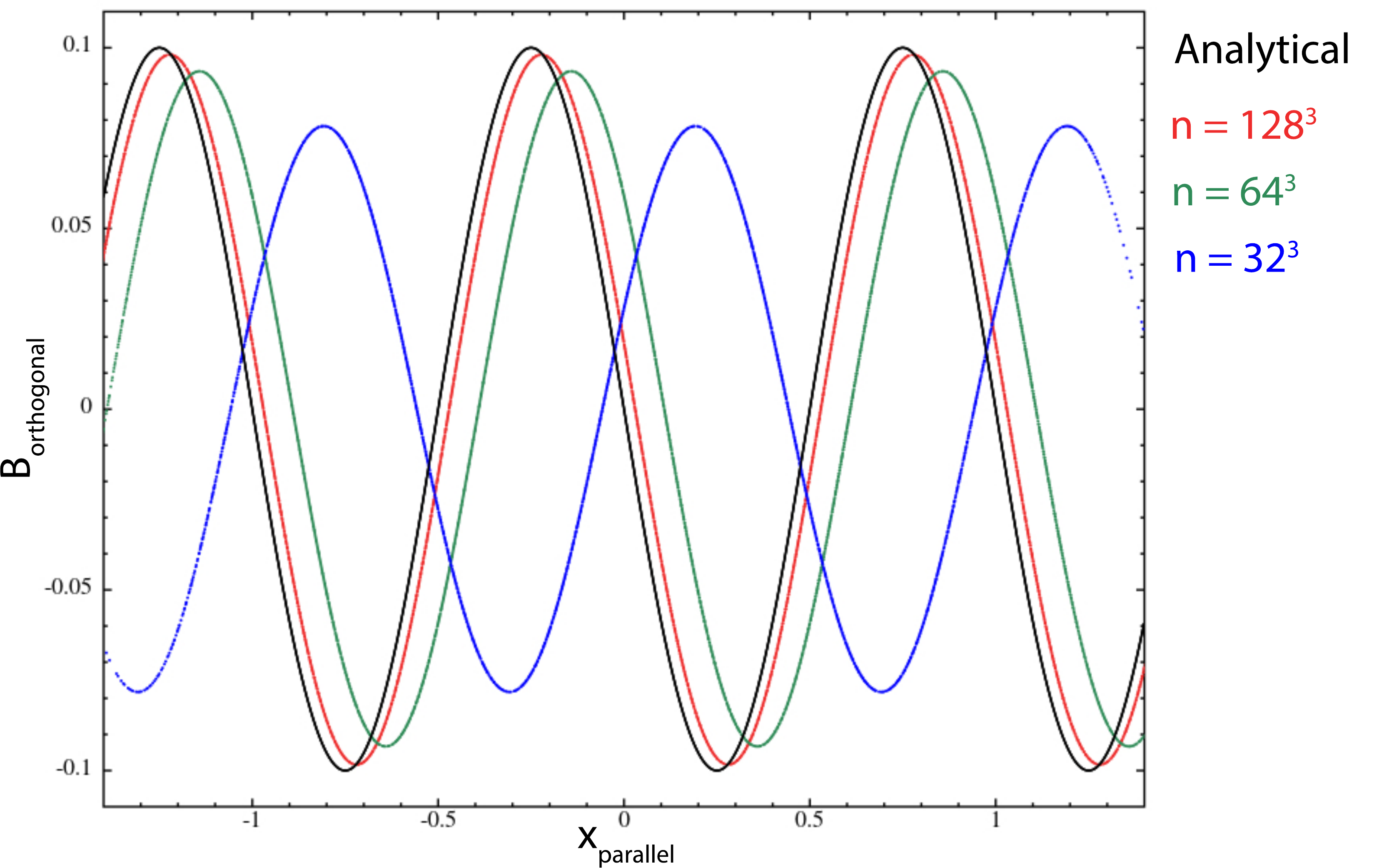}
    \includegraphics[width=\hsize]{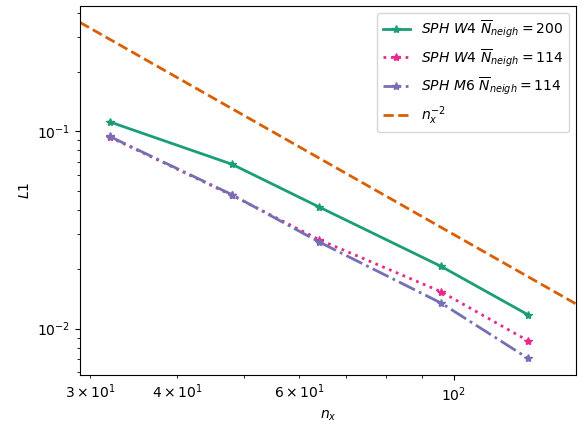}
    \caption{Results of the 3D circularly polarized Alfv\'{e}n wave test. \textbf{Top panel:} transverse component of the magnetic field in the direction of propagation after five periods. The analytical/initial solution is plotted in black, and the simulation results with resolution $[n_x, n_y, n_z] = [128, 74, 78]$ in red, $[n_x, n_y, n_z] = [64, 36, 39]$ in green, and $[n_x, n_y, n_z] = [32, 18, 18]$ in blue. Both of these are with Wendland C4 kernel with 200 neighbours. \textbf{Bottom panel:} convergence study for the Alfv\'{e}n wave test using different kernels and neighbour numbers. Shows how the $L_1$ error scales with resolution (particles along the x-axis). The code default (Wendland C4 kernel with 200 neighbours) is shown in green, Wendland C4 kernel with 114 neighbours are shown in magenta, Quintic kernel with 114 neighbours are shown in blue and the dashed brown line shows the curve for second order convergence. Convergence towards the analytical solution for all kernels are close to second order. When the smoothing length becomes comparable to half the wave length of the Alfv\'{e}n wave, the MHD gradients becomes more ill defined which causes the slower convergence speed for the Wendland kernel ($\overline{N}_{neigh}=200$) at low resolution. }
    \label{fig:alfven}
\end{figure}
The circularized polarized Alfv\'{e}n wave test was introduced by \cite{2000JCoPh.161..605T} to serve as an analytical solution to the ideal MHD equations. Due to the waves being circularized, the gradient in magnetic pressure is zero and the wave should remain the same after each period. This proves to be a useful test for gauging the dissipation and dispersion of the MHD implementation. This test is sensitive to the tensile instability \citep{2005MNRAS.364..384P}, so it also serves as a good test to see if the stability term (\eq~\ref{eq:stabilityterm}) properly stabilizes the solution. The setup follows \cite{2008JCoPh.227.4123G} and \citet{2018PASA...35...31P}, in which the waves are traveling at an angle of $\theta = 30\degree$ with respect to the $x$ axis, within a periodic box of length $L=(l,l/2,l/2)$, where $l = 3$. The transverse velocities and magnetic fields are circularized:
$$B_{\perp,1} = v_{\perp,1} = 0.1\sin{\left(2\pi x_{\parallel}/\lambda\right)} ,$$
$$B_{\perp,2} = v_{\perp,2} = 0.1\cos{\left(2\pi x_{\parallel}/\lambda\right)} ,$$
while the parallel components are set to:
$$v_\parallel=0 \quad B_\parallel=1 .$$
Here, $x_{\parallel}$ is the direction of propagation, and $\lambda=1$ is the wavelength.
An adiabatic EOS ($\gamma=5/3$) is used with uniform pressure $P=0.1$ and density $\rho=1.0$. The particles are set up on a close-packed lattice and the simulation is run for five periods ($t=5$). As we have uniform density, there are no differences between GDSPH and TSPH in this test case. We plot the transverse component of the magnetic field in the direction of propagation, showing the results of different resolutions $[n_x, n_y, n_z] = [128, 74, 78]$ , $[64, 36, 39]$, and $[32, 18, 18]$ in the upper panel of \fig~\ref{fig:alfven}. From the results, we can see that both the phase and amplitude converge towards the analytical solution as we increase resolution. For a more qualitative look at the convergence, we perform a convergence study for this test using different smoothing kernels and neighbour numbers. In the lower panel of \fig~\ref{fig:alfven}, we show the L1 error norm for the transverse magnetic field at five different resolutions($n_x=32,48,64,96,128$), and as we can see all the kernels exhibits second-order convergence. The major outliner is at low resolution for the Wendland C4 kernel with more neighbours. This is caused by the larger smoothing length, which at low resolution becomes comparable to half the wave length of the Alfv\'{e}n wave. This makes the MHD gradients more ill defined which causes force errors that shows itself predominately as a phase shift in the Alfv\'{e}n wave as time goes by. The amplitude of the wave is only weakly affected by this. From the bottom panel in \fig~\ref{fig:alfven} we can also see that the Wendland kernel has slightly lower convergence speed then the quintic kernel at higher resolution. Despite this result, we chose to go with the Wendland kernel C4 with 200 neighbors as our code default for the forthcoming tests. This is for several reasons. First, the quintic kernel is susceptible to the pairing instability whereas the Wendland kernels are not  \citep{2012MNRAS.425.1068D}. In addition, the Wendland kernels tend to make the particle distribution remain well ordered in dynamical conditions, which improves the overall accuracy of the method \citep{2015MNRAS.448.3628R}. While the computational cost increases roughly linearly with increased neighbour number, gravity is usually the more dominant cost in astrophysical simulations, which means that the increase in cost is usually not significant. In the end, the choice of kernel and neighbour number will depend on the application at hand. However, for simulations involving subsonic flows, a high neighbour number has been shown to be preferred (as showcased by the Gresho-Chan vortex test in \cite{2012MNRAS.425.1068D} and \cite{2015MNRAS.448.3628R}).
\subsection{Advection loop}
\label{subsec:advloop}
\begin{figure}[!h]
    \centering
    \includegraphics[width=\hsize]{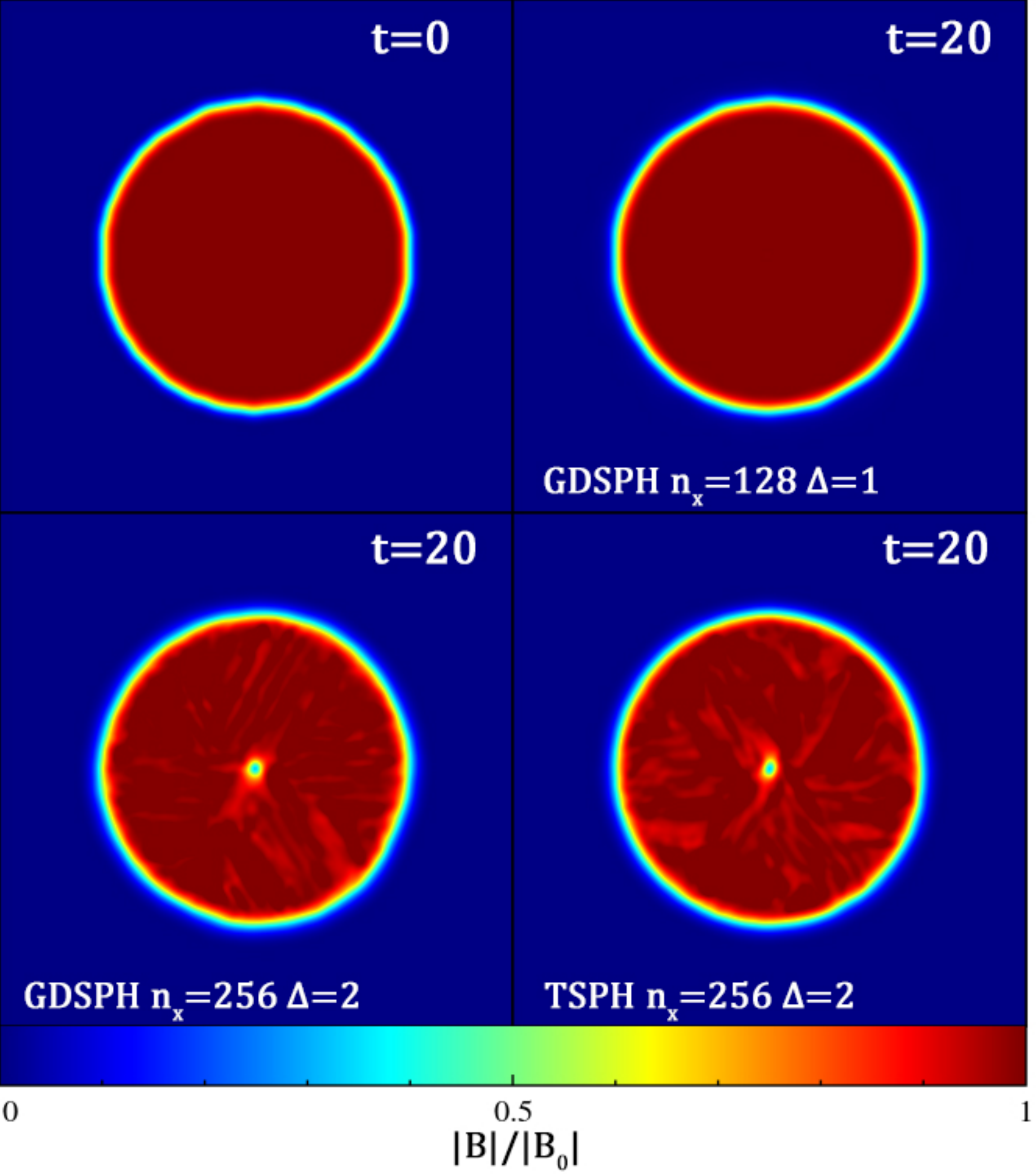}
    \caption{Results from the advection loop test in 3D, showing a rendering of $|\boldsymbol{B}|$ in units of the initial value $|\boldsymbol{B_0}|$. The top left panel shows the initial setup (same rendering for $n_x=128$ and $n_x=256$, and the top right shows the uniform density case for GDSPH with resolution $n_x=128$ after twenty crossings $t=20$. We can see that the current loop is conserved almost perfectly even with all the dissipation terms turned on. This shows an improvement comparing to the SPMHD results in {\sc Phantom}, where dissipation is seen after five periods \citep[Figure 35 in][]{2018PASA...35...31P}, which plots the current density. The bottom two panels show the cases with a density gradient $\Delta \equiv \frac{\rho_{in}}{\rho_{out}}=2$ between the inner loop and outer medium and with resolution of $n_x=256$, after twenty crossings $t=20$.  The bottom left panel shows the GDSPH case and the bottom right show the TSPH case ($n_x=256$). Both show similar dissipation of the magnetic field to the results from the MFM/MFV method in {\sc Gizmo} \citep{2016MNRAS.455...51H}.}
    \label{fig:loop}
\end{figure}
\begin{figure}[!h]
    \centering
    \includegraphics[width=\hsize]{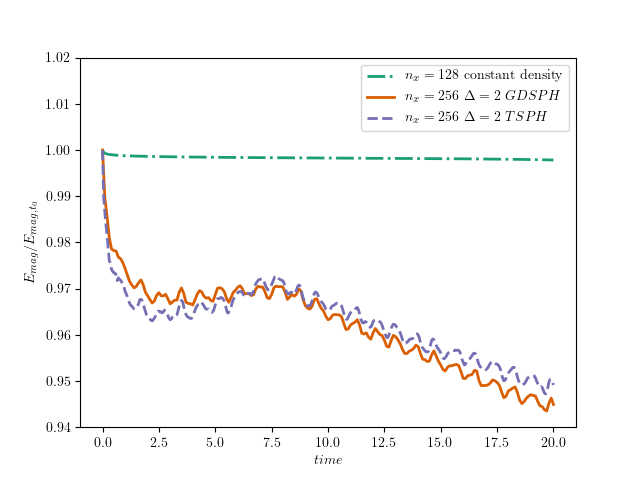}
    \caption{Results from the advection loop test in 3D, showing the time evolution of the magnetic field energy in units of the initial value. After $t=20$ our uniform case has dissipated about $0.3\%$, while the cases with the density gradient $\Delta=2$ have dropped around $5\%$, which is similar to the dissipation displayed by the MFM/MFV method in \cite{2016MNRAS.455...51H}. There is a tiny difference between GDSPH and TSPH, but this owes itself to differences in the initial reordering.}
    \label{fig:loopE}
\end{figure}
The advecting current loop test was introduced by \cite{2005JCoPh.205..509G,2008JCoPh.227.4123G}, in which a weak magnetic loop is advected by a constant velocity field. As the ratio between the thermal pressure and magnetic pressure is massive ($\beta\approx 10^6$), the magnetic field is dynamically unimportant and should simply be advected along the velocity field. This proves to be one of the more difficult tests for grid-based code due to intrinsic dissipation during advection. We followed the setup from \citet{2008JCoPh.227.4123G}, \citet{2016MNRAS.455...51H} and \citet{2018PASA...35...31P}, and initialized a 3D thin periodic box with length $L=(2,1,2\frac{\sqrt{6}}{n_x})$, velocity $v=(2,1,\frac{0.1}{\sqrt{5}})$ and pressure $P=1$. The magnetic field inside the loop was determined from the potential $A_z=A_0(R_0-r)$, where $A_0=10^{-3}, \ R_0=0.3$, and $r^2=x^2+y^2$. The face-centered magnetic fields are then $\boldsymbol{B_0}=\nabla \times A_z = \frac{A_0}{r}(y,-x,0)$ inside the loop and zero everywhere else. We set up two initial conditions, one with a uniform density $\rho=1$ with resolution $[n_x, n_y, n_z] = [128, 74, 12]$ and another with a density gradient ($\Delta \equiv \frac{\rho_{in}}{\rho_{out}}=2$) between the inner loop ($\rho_{in}=2$) and outer medium ($\rho_{out}=1$), with resolution $[n_x, n_y, n_z] = [256, 148, 12]$. The particles are set up on a close-packed lattice and the loop is advected for twenty periods with all the default dissipation and divergence cleaning terms turned on. The results of the test can be seen in \fig~\ref{fig:loop} and \fig~\ref{fig:loopE}. 
\\ \\
From the figures, we can see that in the case with uniform density the field loop is closely conserved, resulting in only $0.3 \%$ reduction in magnetic energy after twenty periods. This is a significant improvement from the advection loop presented in \cite{2018PASA...35...31P}, which starts to degrade after five periods. We have also tested this case using a quintic spline kernel, a smaller number of neighbours and an $\alpha_B=1$ for the AR similar to \cite{2018PASA...35...31P}, while a little more degradation can be seen, the difference in magnetic energy is still small after twenty periods ($1\%$ instead of a $0.3\%$ decrease in magnetic energy). Without any dissipation and divergence cleaning the advection loop with uniform density can be sustained for thousands of periods, as shown in \cite{2007MNRAS.379..915R}. This shows a significant advantage for Lagrangian codes compared to Eulerian codes, which suffer from resolution-dependent advection errors when the configuration is not aligned to the grid. In the case of the density gradient, we can see that there is now a faster dissipation of the magnetic energy. The sudden reduction in magnetic energy is largely due to the reordering of the initial particle lattice near the density contrast. Comparing to \cite{2016MNRAS.455...51H}, we can see that we have a similar reduction in magnetic energy as in the results from the MFM/MFV method at $t=20$. There is a tiny difference ($< 1\%$) between GDSPH and TSPH, however, this owes itself to the initial reordering, after that we can see that the rate of change in the two discretizations are practically the same. The averaged normalized divergence error, $\langle\epsilon_{divB}\rangle$, is around $10^{-2}$ for both the $\Delta = 1$ case and $\Delta = 2$ case.
\subsection{Brio-Wu shocktube}
\label{subsec:brio}
\begin{figure*}[!hp]
    \centering
    \includegraphics[width=\hsize]{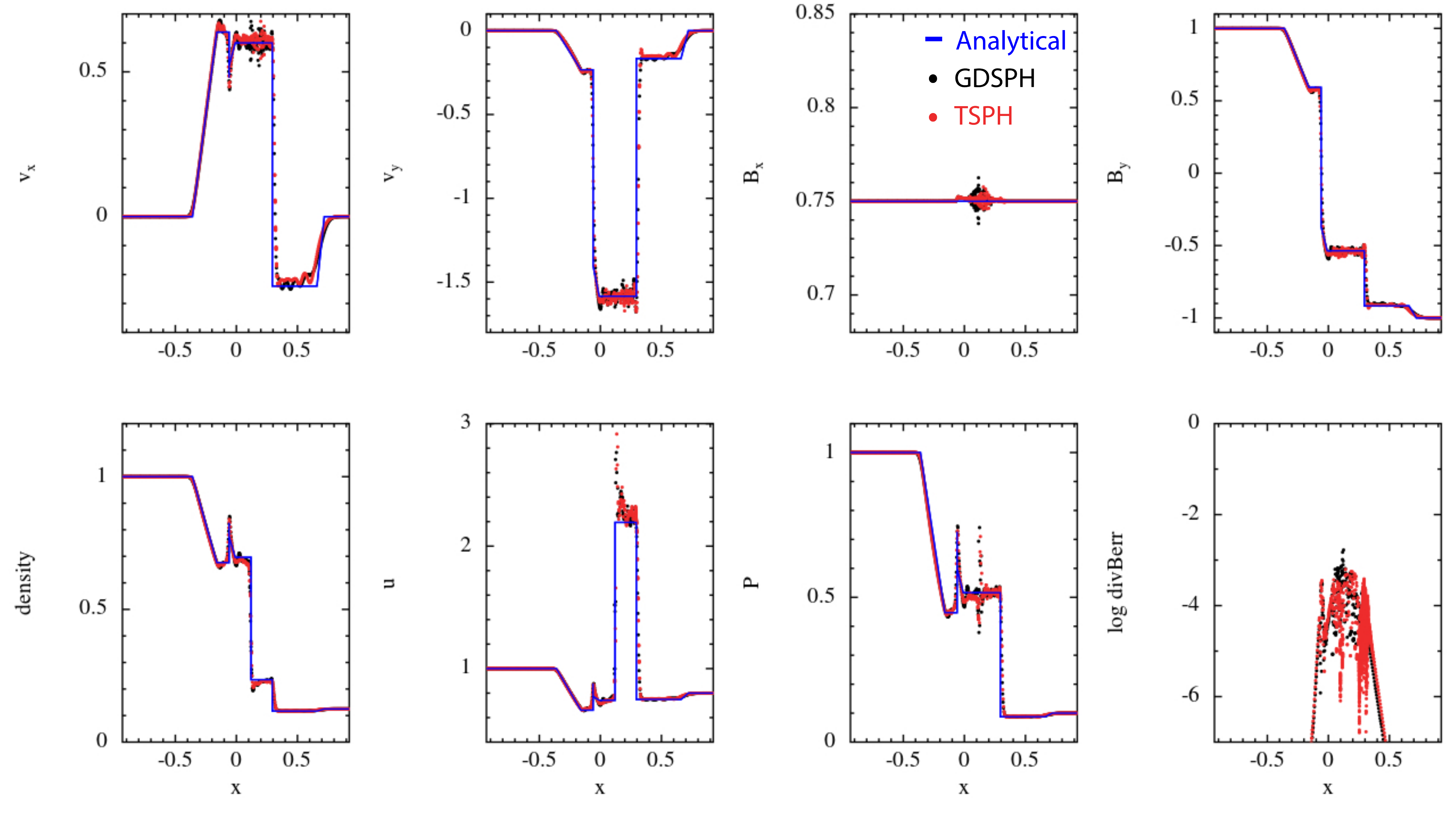}
    \caption{Results from the Brio-Wu shocktube in 3D, with an initial left state $(\rho_L,P_L,v_{x},v_y,v_z,B_x,B_y,B_z)=(1,1,0,0,0,0.75,1,0)$ and right state $(\rho_R,P_R,v_x,v_y,v_z,B_x,B_y,B_z)=(0.125,0.1,0,0,0,0.75,-1,0)$. The figure shows the active region of the shock after $t=0.2$, which contains about $n_x\approx300-400$ particles across the x-direction. The blue line shows the reference solution and the black dots show the result from the GDSPH simulation, while red dots show the result from the TSPH simulation. There are minimal differences between the GDSPH and TSPH result.}
    \label{fig:brio}
\end{figure*}

The Brio-Wu shocktube \citep{1988JCoPh..75..400B} is an MHD extension to the classic Sod shocktube test (the hydro setup is the same). It tests how well the implementation can handle different MHD shocks, rarefactions, and contact discontinuities. We followed the setup from \citet{2016MNRAS.455...51H} and \citet{2018PASA...35...31P} and initialized a 3D thin periodic box, with a total region of $[n_x, n_y, n_z] = [1024, 24, 24]$ and an active region of $[256, 24, 24]$ particles on the left side $x_L=[-2,0]$, and a total region of $[n_x, n_y, n_z] = [512, 12, 12]$ and an active region of $[128, 12, 12]$ on the right side $x_R=[0,2]$.
The left state was set to:
$$(\rho_L,P_L,v_{x},v_y,v_z,B_x,B_y,B_z)=(1,1,0,0,0,0.75,1,0) ,$$
and the right state was set to:
$$(\rho_R,P_R,v_x,v_y,v_z,B_x,B_y,B_z)=(0.125,0.1,0,0,0,0.75,-1,0) .$$
The adiabatic index is set to $\gamma=2$. We ran the simulation up to t=0.2 and the results can be seen in \fig~\ref{fig:brio}. The GDSPH and TSPH results are shown with black and red dots, respectively, and the analytical solution is shown in blue lines. 
Our results are very similar to those from the {\sc Phantom} code default case with the same resolution \citep[Figure 30 in][]{2018PASA...35...31P}. However, there is noticeable wall heating in the internal energy $u$ in our test, due to the conservative thermal dissipation term used in this work. A more aggressive thermal dissipation can be added to smooth out the wall heating, which improves the results in the density, thermal and pressure profiles. However, this often leads to over dissipation in cases involving gravitational fields and is thus not a preferable choice. Divergence errors are kept low with a maximum value of $\sim 10^{-3}$ at the shock, and $B_x$ remains close to constant, which also indicates excellent divergence control. Varying artificial resistivity parameter $\alpha_B$ from $\alpha_B=0.5$ to $\alpha_B=1 $ only shows minimal differences, and as we can see from the results, $\alpha_B=0.5$ is sufficient to capture the magnetic field structure. We also note that using constant artificial viscosity (AV) parameters decreases post-shock oscillations and improves the results in the velocity profile. From \fig~\ref{fig:brio}, we can see that there is very little difference between GDSPH and TSPH in this test. 
\subsection{Orszag-Tang vortex}
\label{subsec:orz}
\begin{figure*}[!hp]
    \centering
    \includegraphics[width=12cm]{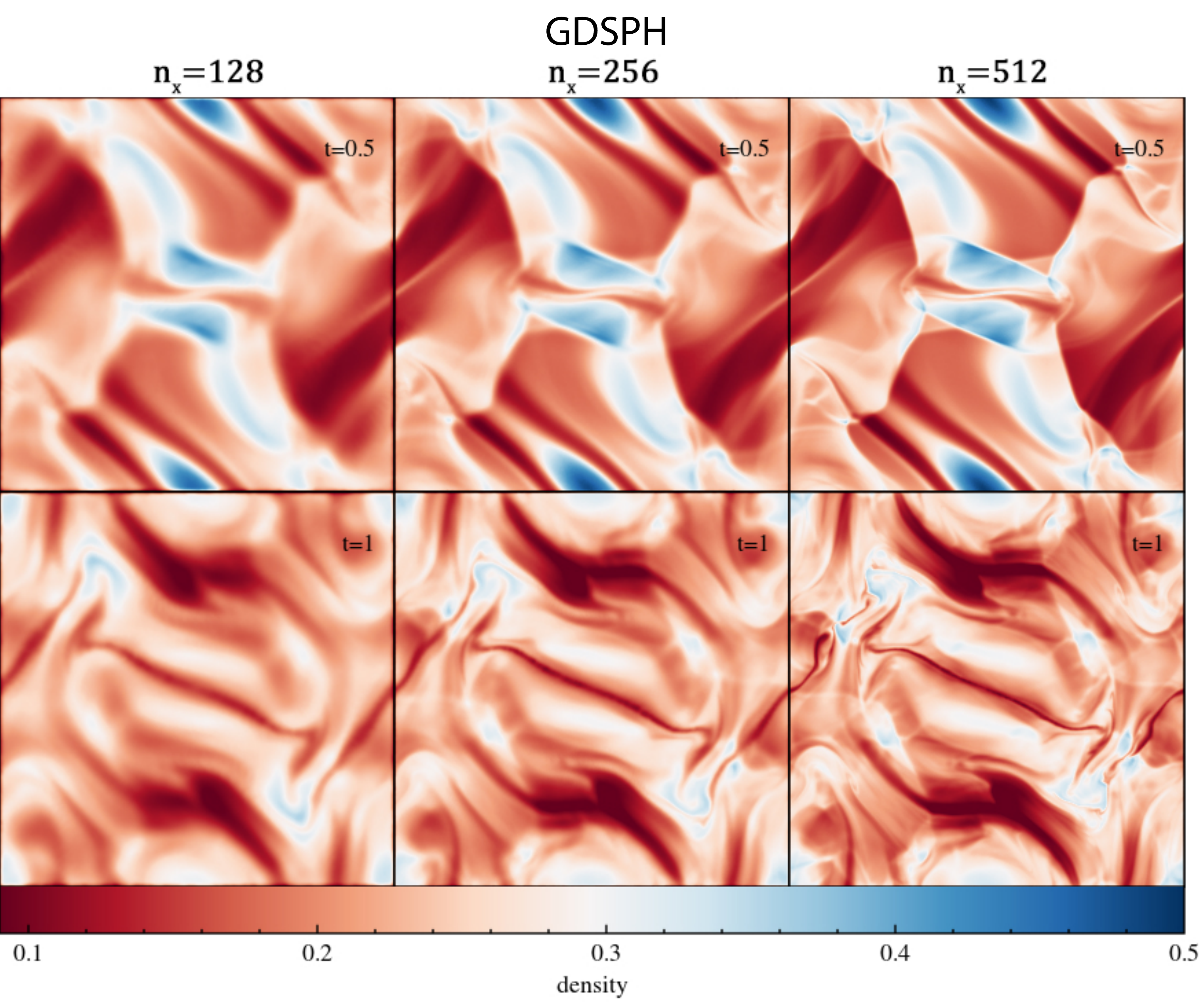}
    \caption{Results from the Orszag-Tang vortex in 3D done with GDSPH, which shows rendered density slices ($z=0$) at $t=0.5$ (top) and $t=1$ (bottom) for varying resolution $[n_x,n_y,n_z]=[128,148,12],[256,296,12]$ and $[512,590,12]$ (low to high from left to right). The simulations capture well most of the key features for all tested resolutions. With increasing resolution the flows are more defined and show increased complexity. There are no significant differences between GDSPH and TSPH in this case.}
    \label{fig:orz}
\end{figure*}

\begin{figure}[h!]
    \centering
    \includegraphics[width=\hsize]{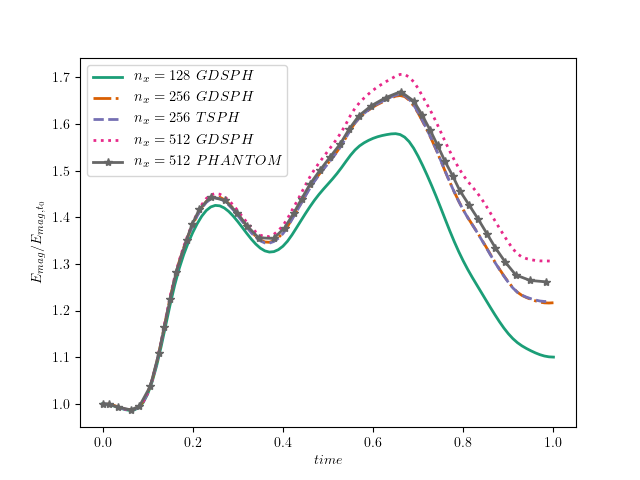}
    \caption{Evolution of the total magnetic energy in units of the initial value for the 3D Orszag-Tang vortex test. We plot the result from three different tested resolutions $[n_x,n_y,n_z]=[128,148,12],[256,296,12],[512,590,12]$ done in GDSPH. We also include a TSPH case with $n_x=256$ and the $n_x=512$ curve from \cite{2017arXiv170607721W} for comparison sake. We can see that there are no visible difference between the TSPH (purple curve) and GDSPH (brown curve) cases. From the figure, it is also clear that the GDSPH case of $n_x=512$ (magenta curve) is less dissipative than the simulation from \cite{2017arXiv170607721W} (grey curve). Significant differences between resolutions can be seen to occur at later times during the evolution.}
    \label{fig:orzE}
\end{figure}

The Orszag-Tang vortex test was introduced by \cite{1979JFM....90..129O} and is a standard test of MHD schemes, as it involves the development of super-sonic turbulence and the interaction of the different MHD shocks. We set up a 3D thin periodic box with $L=(1.0,1.0,\frac{2\sqrt{6}}{n_x})$ at varying resolutions ($[n_x,n_y,n_z]=[128,148,12], [256,296,12]$ and $[512,590,12]$). The test consists out of a velocity vortex: $$[v_x,v_y,v_z]=v_0[-\sin\left(2\pi (y-y_{min})\right),\sin\left(2\pi (x-x_{min})\right),0] ,$$ and a doubly periodic magnetic field:
$$[B_x,B_y,B_z]=B_0[-\sin\left(2\pi (y-y_{min})\right),\sin\left(4\pi (x-x_{min})\right),0] ,$$
where $v_0=1$, $B_0=1/\sqrt{4\pi}, x_{min}=-0.5$ and $y_{min}=-0.5$. Setting the initial plasma beta $\beta_0=10/3$, the initial Mach number $M=v_0/c_s=1$ and the adiabatic index $\gamma=5/3$, we get the initial pressure $P_0=B_0^2\, \beta_0 / 2 = 0.133$ and density $\rho_0=\gamma P_0 M_0 = 0.221$. We show the results of the different resolution runs in \fig~\ref{fig:orz} after t=0.5 (top row) and t=1 (bottom row). The test was run with both GDSPH and TSPH, however, we found only very small differences between them, which is why we only show the result from GDSPH. This can additionally be seen in \fig~\ref{fig:orzE}, which show the time evolution of the magnetic energy in all of our test cases. From the result at t=0.5 we can see that we reproduce the shock structure well and capture the trapped dense filament in the centre of the domain for all resolutions. With increasing resolution, the shock structure and filament become more defined. At t=1, a more turbulent flow has developed. Our simulations capture most of the key features and compare well with previous works (for example, Figure 32 in \citet{2018PASA...35...31P}). However, it appears that our method is unable to develop the central magnetic island, a feature that is supposed to form when the current sheet in the center becomes unstable and reconnects due to the tearing mode instability. This is the case for most previous implementations of SPMHD, however, in \cite{2017arXiv170607721W} the authors argue that with a less dissipative artificial resistivity switch the magnetic island can be reproduced, and this motivate them to use the signal speed $v_{sig, B}$ given in \eq~\ref{eq:vsig}.
\begin{figure}[]
    \centering
    \includegraphics[width=\hsize]{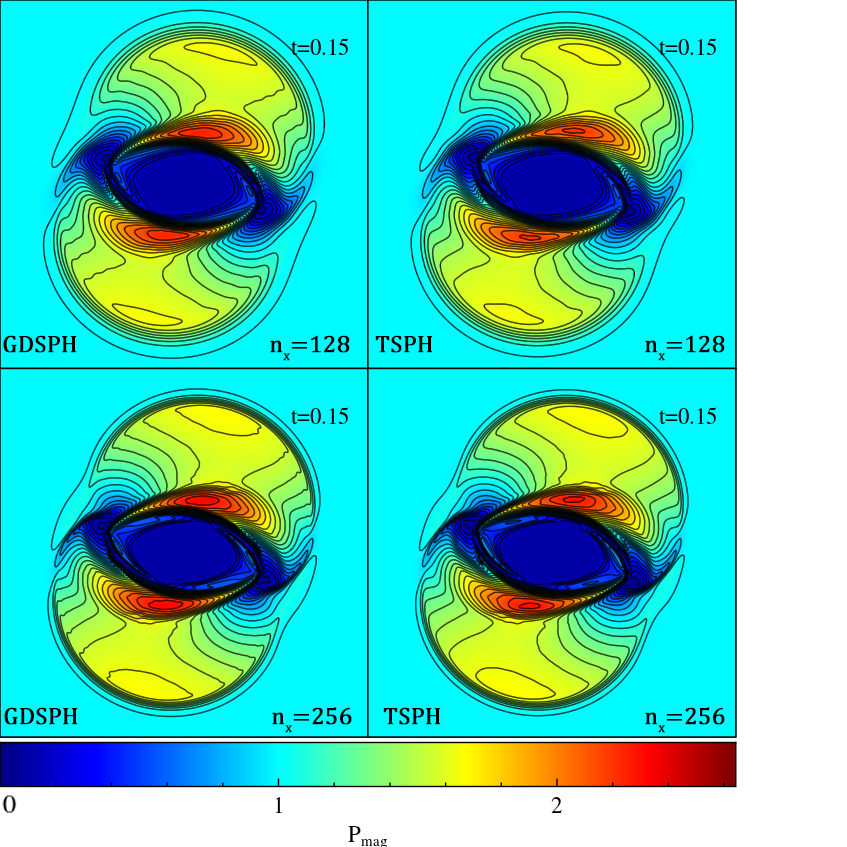}
    \caption{Result from the magnetic rotor in 3D, which shows rendered magnetic pressure slices ($z=0$) at $t=0.15$, for varying resolution $[n_x,n_y,n_z]=[128,148,12]$ and $[256,296,12]$ and for both GDSPH and TSPH. The plot also shows 30 contours with limits taken to be the same as \cite{2000JCoPh.161..605T} and \cite{2018PASA...35...31P} $(P_{mag}=[0,2.642])$ for a more direct comparison. We can see that GDSPH develops a larger and broader magnetic pressure peak compared to TSPH. }
    \label{fig:rotor}
\end{figure}
We use the same switch but the problem remains. Nevertheless, if we compare our evolution of the magnetic energy (the magenta curve in \fig~\ref{fig:orzE}) to theirs (the grey curve) we can see that our simulations are actually less dissipative, likely because we use a smaller resistivity parameter $\alpha_B=0.5$.  Increasing $\alpha_B $ to 1 leads to more dissipation and brings the final energy closer to the Phantom run. It is thus likely that the development of the magnetic island also depends on the other dissipation terms such as AV and artificial conductivity. The mean normalized divergence error in the simulations are of the order $\langle\epsilon_{divB}\rangle = 10^{-3.5}$ to $10^{-2.5}$, decreasing with higher resolution.
\subsection{MHD rotor}
\label{subsec:rotor}
The MHD rotor test was introduced by \cite{1999JCoPh.149..270B}, which tests the propagation of Alfv\'{e}n waves generated by a magnetized rotor. We followed the setup from \cite{2000JCoPh.161..605T} and \cite{2018PASA...35...31P}, and initialized a 3D thin periodic box of $L=(1.0,1.0,\frac{2\sqrt{6}}{n_x})$ at two resolutions, $[n_x,n_y,n_z]=[128,148,12]$ and $[256,296,12]$. A rotating dense cylinder ($\rho=10$) was initiated with cylindrical radius $R=0.1$, within a surrounding medium ($\rho=1$). The cylinder was put into rotation with an initial velocity of $$v=\frac{v_0}{r_{cyl}}[-(y-y_0),(x-x_0),0] \quad r_{cyl}<R ,$$ where $r_{cyl}=\sqrt{x^2+y^2}$ and $v_0=2$. The initial pressure was set to $P=1$, with an adiabatic index of $\gamma=1.4$. The initial magnetic field was set to $B=[5/\sqrt{4},0,0]$. The particles were set up on a closed packed lattice and the simulation were run until $t=0.15$. The density contrast was unsmoothed, which means that there will be some noise at the boundary initially, due to particle reordering. The results of the simulations done with GDSPH and TSPH can be seen in \fig~\ref{fig:rotor}, which shows 30 contours and the rendering of the magnetic pressure, with limits taken to be the same as in \cite{2000JCoPh.161..605T} and \citet{2018PASA...35...31P}. From the results, we can see that the difference between GDSPH and TSPH is generally small.  However, we do find that in GDSPH there are notable increases in magnetic pressure at the pressure maxima compared to TSPH, which also seen in \cite{2016MNRAS.455...51H} when the authors compared MFM with SPH (their Figure 15). In general, the key features of the test are captured by both methods. The mean normalized divergence errors in the simulations are of the order $\langle\epsilon_{divB}\rangle = 10^{-4}$ to $10^{-3}$.
\begin{figure}[!hp]
    \centering
    \includegraphics[width=\hsize]{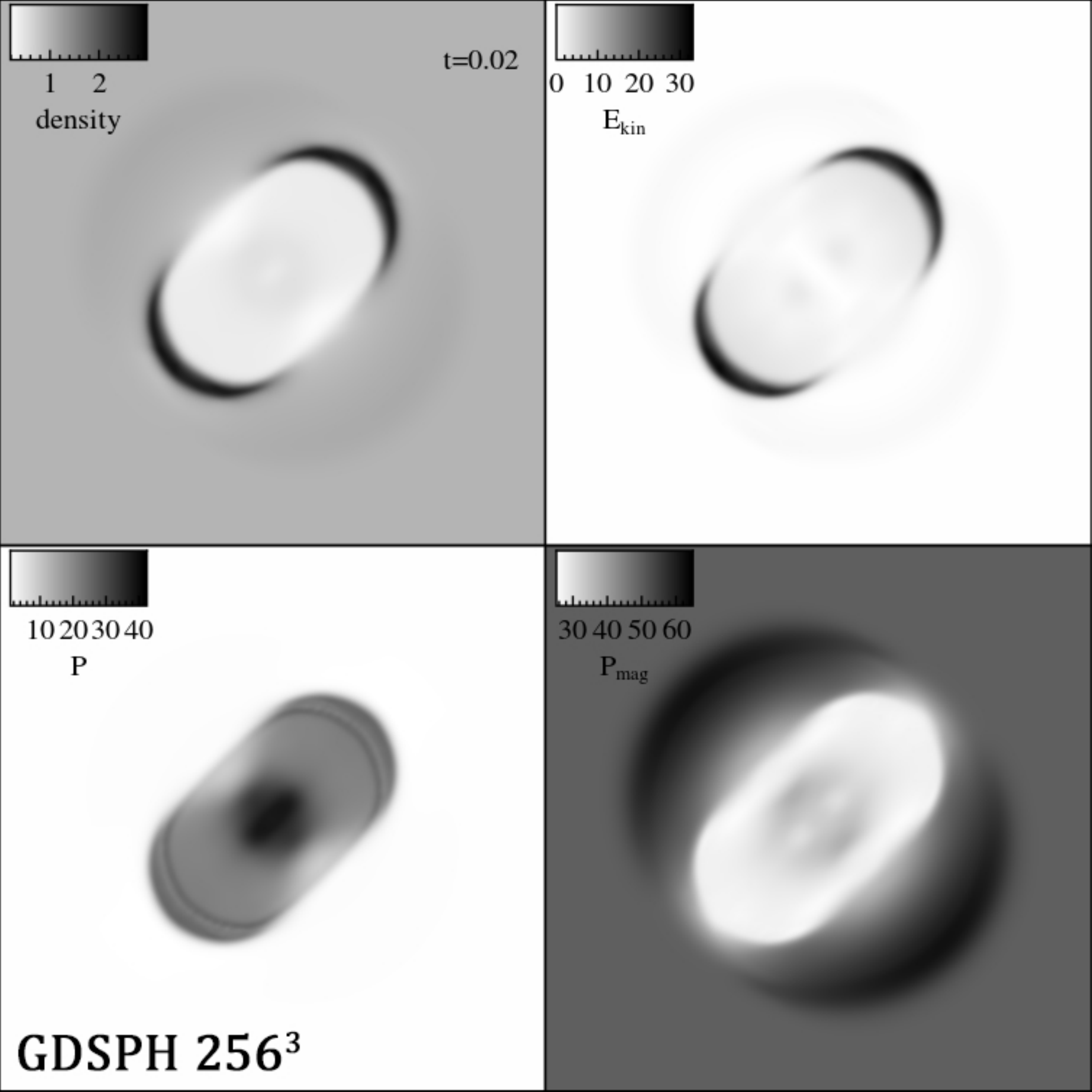}
    \caption{Result from the magnetized blast in 3D in the $256^3$ GDSPH run, which shows rendered slices of different fluid quantities at $t=0.02$. To the top left we can see the density rendering, top right the kinetic energy density, bottom left the thermal pressure and bottom right the magnetic pressure. The limits are taken to be the same as \cite{2008ApJS..178..137S} and \citet{2018PASA...35...31P} for a direct comparison: $\rho=[0.19,2.98]$, $E_{kin}=[0,33.1]$, $P=[1,42.4]$ and $P_{mag}=[25.2,65.9]$}
    \label{fig:blast}
\end{figure}
\begin{figure}[!h]
    \centering
    \includegraphics[width=\hsize]{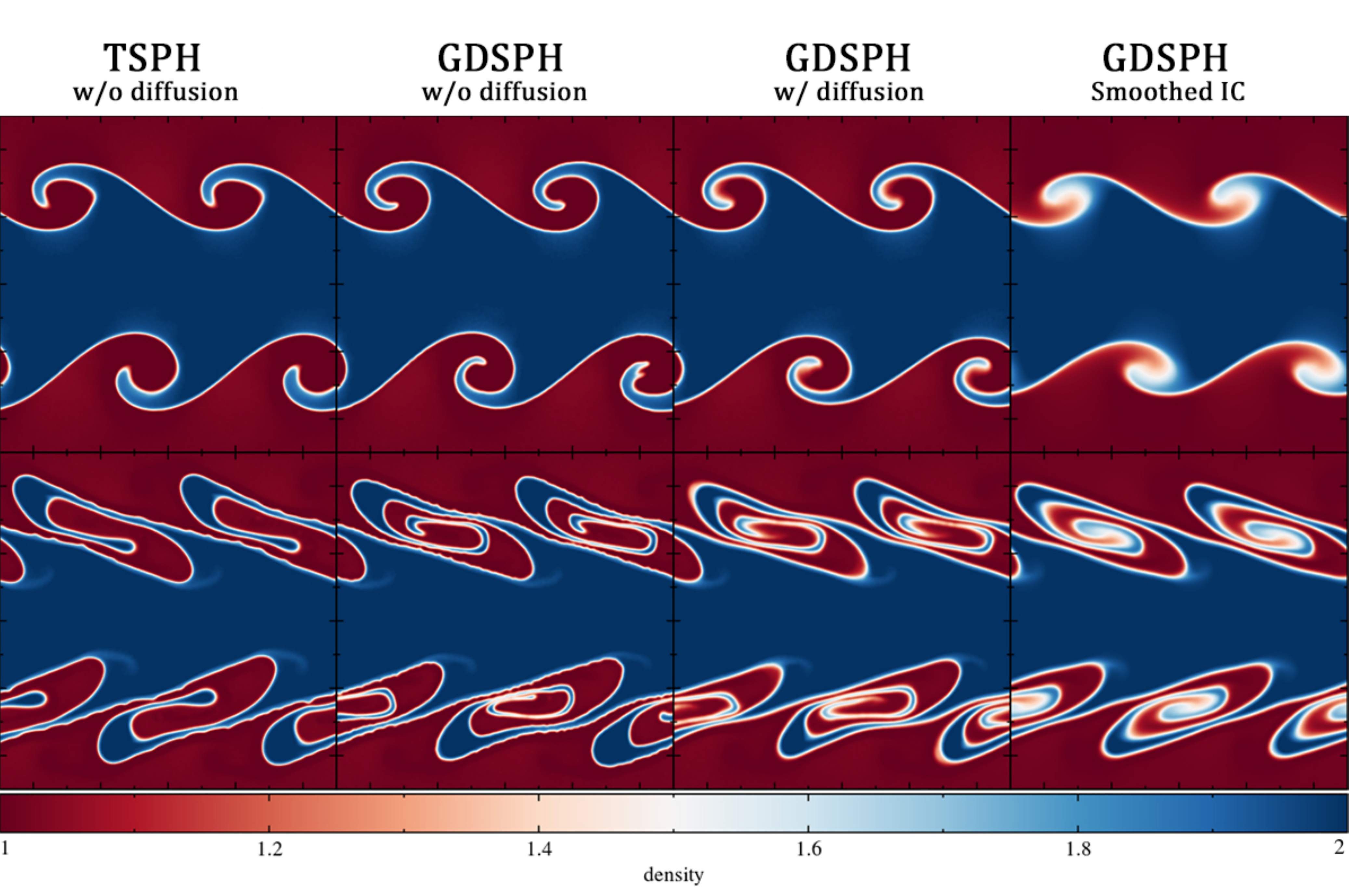}
    \caption{Result from the magnetic Kelvin-Helmholtz instability in 3D, which shows rendered density slices ($z=0$) at $t=1.6$ (top) and $t=3.2$ (bottom), for TSPH without diffusion (left), GDSPH without diffusion (middle left), GDSPH with diffusion (middle right) and GDSPH with an initial smoothed contact discontinuity (right). The TSPH result exhibits a very gloopy behaviour and shows decreased growth of the KH mode. This is mainly due to the artificial surface tension effect (see \fig~\ref{fig:khpart}). GDSPH shows much better growth, the addition of diffusion only slightly improves the growth rate. The main effect from the magnetic field can be seen in all cases, which uncoils and stretches the vortex. The smoothed result closely resembles the MFM and grid result from \cite{2016MNRAS.455...51H}}
    \label{fig:kh}
\end{figure}
\subsection{Magnetized blastwave}
\label{subsec:mhdblast}
The magnetized blastwave was introduced by \cite{1999JCoPh.149..270B} and \cite{2000ApJ...530..508L}, in which a central over-pressurized region expands preferentially along the magnetic fields lines. We followed the setup from \cite{2008ApJS..178..137S} and \cite{2018PASA...35...31P}, and initialized a 3D periodic box of $L=[1.0,1.0,1.0]$ with uniform density at a resolution of $N=256^3$. An inner region of radius $R=0.125$ was over-pressurized ($P_{in}=100$) to 100 times the outer pressure, which was set to $P_{out}=1$. The adiabatic index of the gas is set to $\gamma=5/3$.\footnote{This is different from the choice in \cite{2018PASA...35...31P} ($\gamma=1.4$), however, $\gamma=5/3$ is more representative of gas in astrophysical applications} The initial magnetic field was set to:
$$B=[10/\sqrt{2},0,10/\sqrt{2}] .$$
This sets the initial plasma beta to $\beta_{in}=2$ in the inner region and $\beta_{out}=0.02$ in the outer region. The simulation was run for $t=0.02$ and the result can be seen in \fig~\ref{fig:blast}. The rendering and limits in the figure were set to the same as the results presented in \cite{2000JCoPh.161..605T} and \citet{2018PASA...35...31P}, so that they can be directly compared. We can see that our results agree well with the previous authors, capturing the inner and outer structure of the blast well. There are minimal differences between the GDSPH and TSPH results. The mean normalized divergence error in the simulations are of the order $\langle\epsilon_{divB}\rangle = 10^{-5}$.
\begin{figure*}[!h]
    \centering
    \includegraphics[width=\hsize]{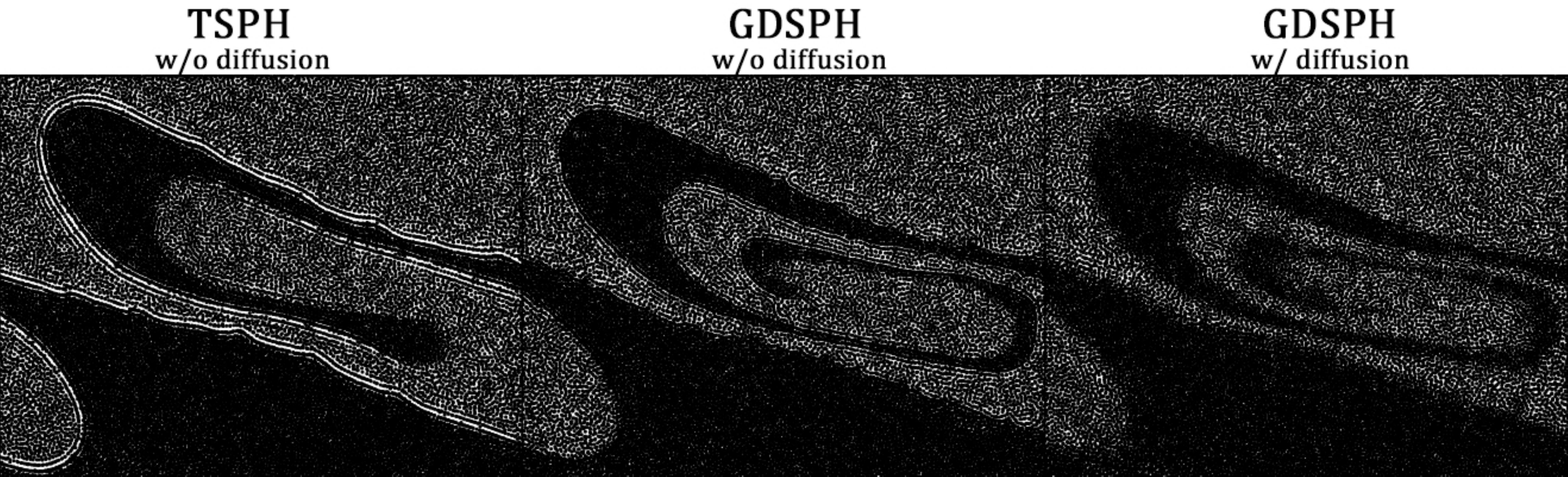}
    \caption{Surface boundary between the high and low density region in the magnetic Kelvin-Helmholtz instability at t=3.2. The effect of the numerical surface tension can clearly be seen in the TSPH case, while GDSPH does not suffer from this issue. Adding thermal diffusion allows for local mixing between the cold and hot phase.}
    \label{fig:khpart}
\end{figure*}
\subsection{Kelvin-Helmholtz instability in MHD}
\label{subsec:kh}

Generation of the Kelvin-Helmholtz instability is crucial for efficient mixing in hydrodynamical simulations. Here, we look at the same problem but with a magnetic field applied parallel to the flow. This has a stabilizing effect on the shear flow due to the magnetic tension force. We followed the setup from \cite{2012ApJS..201...18M}, but extend it to 3D, making a thin periodic box ($L=[1.0,1.0,\frac{2\sqrt{6}}{n_x}]$), with a resolution of $[n_x,n_y,n_z]=[256,296,12]$. We applied a uniform pressure of $P=5/2$ with an adiabatic index of $\gamma=5/3$. The hot outer stream has a density of $\rho_{out}=1$ and velocity $v_{out}=[-0.5,0,0]$. The cold inner stream has a density of $\rho_{in}=2$ and velocity $v_{in}=[0.5,0,0]$. A uniform magnetic field was added in the direction of the flow velocity $B=[0.1,0,0]$. 
\\ \\
The results for TSPH and GDSPH at $t=1.6$ and $3.2$ can be seen in \fig~\ref{fig:kh}. And in \fig~\ref{fig:khpart} we show the particle distribution of the surface boundary at $t=3.2$. The TSPH result exhibits a very gloopy behaviour and shows a decreased growth of the KH mode. A strong artificial surface tension effect can clearly be seen between the hot and the cold phase in \fig~\ref{fig:khpart}. With GDSPH this effect is largely eliminated and the growth rate improves significantly. This large improvement in GDSPH lends itself mainly to the reduction of the leading order errors, which we discussed in Section~\ref{subsec:sphdisc}. Adding turbulent diffusion (that is, with the code default parameters) further improves the result, because it allows particles to effectively mix or reorder (as shown clearly in the particle distribution near the boundary regions in \fig~\ref{fig:khpart}), but the growth rate remains similar to the GDSPH-only case. We note that the sharp contact discontinuities in the initial condition are not smoothed, unlike in \cite{2012ApJS..201...18M}. We choose this because it represents an extreme situation for SPH where the initial particle ordering is not optimal (that is, zeroth order errors are relatively high), and we show that GDSPH performs well even in this extreme case.
We also ran the setup using a smoothed contact discontinuity, and this is shown in the rightmost column in \fig~\ref{fig:kh}. The magnetic field effectively uncoils and stretches the vortex, which is in good agreement to the results shown in \cite{2016MNRAS.455...51H} with the same setup. Here TSPH and GDSPH develop indistinguishably until later time, where at the end only small differences can be seen. The mean normalized divergence error in the simulations are of the order of $\langle\epsilon_{divB}\rangle = 10^{-3}$.

\section{Collapse of a magnetized cloud}
\label{sec:collapse}
    In this Section, we apply our method to an astrophysical problem and consider the collapse of a magnetized cloud. In this type of problem involving large dynamic scales, we see a substantial difference between GDSPH and TSPH. A rotating magnetized cloud is allowed to collapse under its own gravity. During the collapse, the cloud is compressed over several orders of magnitude, testing how the magnetic field develops and interacts with the gas during compression. The large-scale collapse is eventually halted by the formation of a pseudo-disk\footnote{The disks formed in strong magnetic fields are primarily not supported by rotation, as magnetic braking quickly transfers angular momentum outwards. A pseudo-disk is however formed, which structure is a consequence of the anisotropy of the magnetic support against the gravitational collapse. Due to our high initial rotation rate, we are though more likely to retain a more rotationally supported disk compared to other studies which apply a lower ratio.}, which then starts to slowly contract via magnetic braking. The collapse continues within the central region and as the first hydrostatic core starts to form, the magnetic field is twisted until it eventually launches a jet \citep{1986CaJPh..64..507U,1996MNRAS.279..389L,2000ApJ...541L..21U,2004ApJ...617..123N}. The formation and subsequent evolution of the first hydrostatic core stalls the collapse and a slow contraction phase begins. In this paper, we do not run the simulations far beyond the time of jet launching. The two main jet launching mechanisms are the magneto-centrifugal and the magnetic pressure driven mechanism. With a global poloidal magnetic field as in our model, both of these mechanisms play an important role. The resulting magnetic field structure of the jet consists of a poloidal dominated central core with a surrounding toroidal field which produces a strong current along the jet. We refer to this magnetic field structure as the magnetic tower throughout this paper. All these key aspects require the code to have excellent angular momentum conservation, small numerical dissipation and maintain low divergence errors ($\nabla \cdot B$). 
    \\ \\
\begin{figure*}[]
   \centering
\begin{tabular}{c@{}c}
\includegraphics[width=17cm]{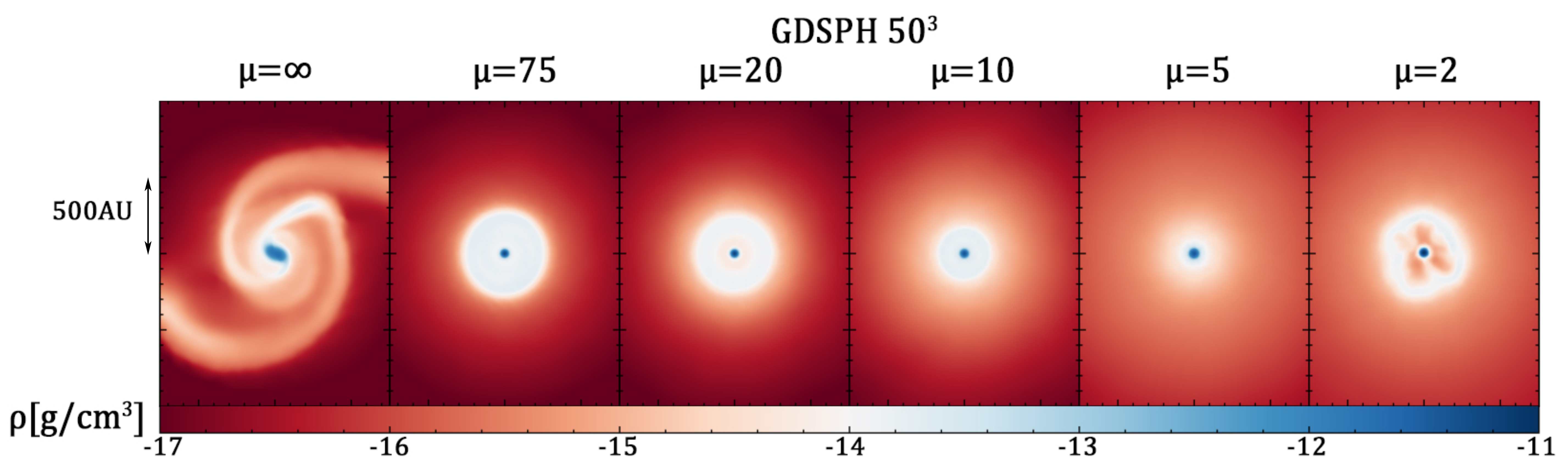}\\
\includegraphics[width=17cm]{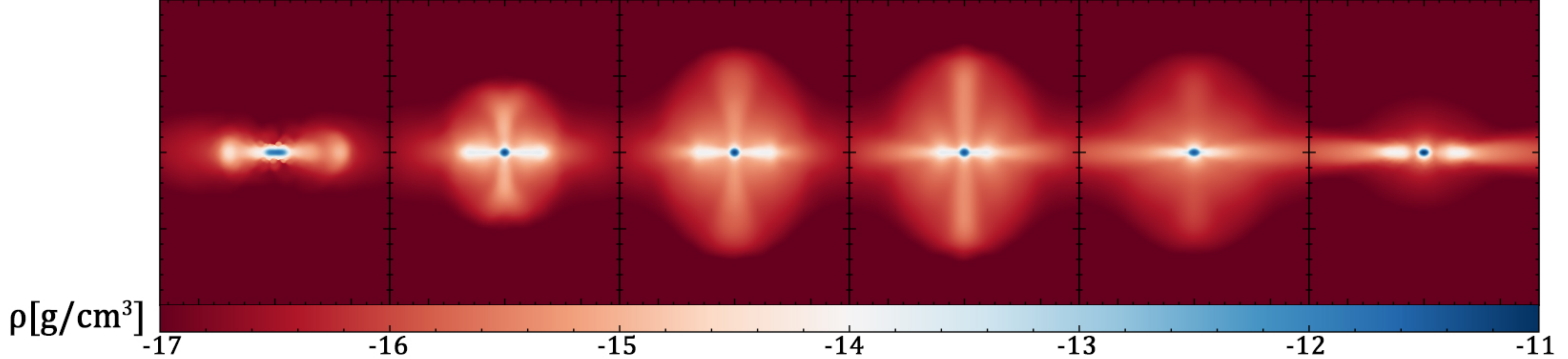}\\
\includegraphics[width=17cm]{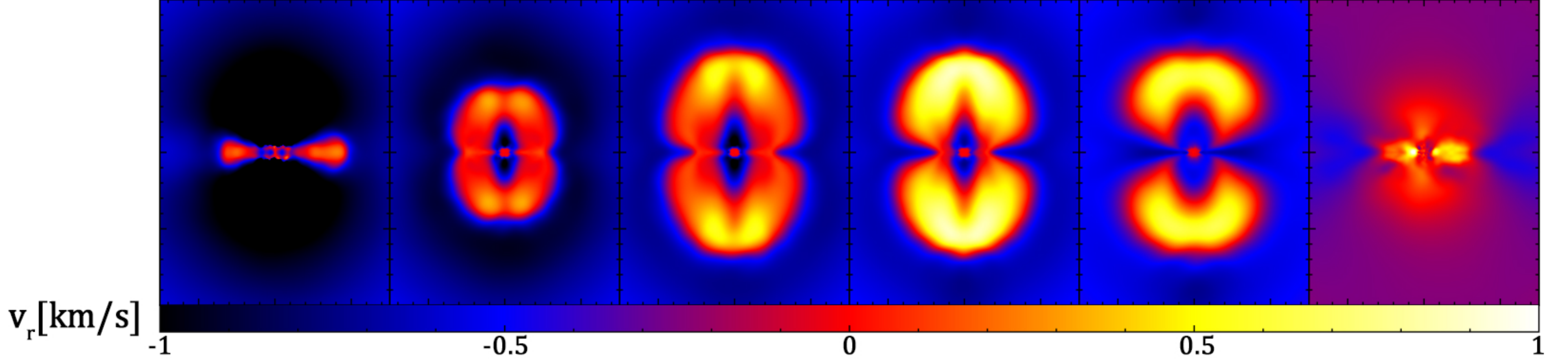}\\
\includegraphics[width=17cm]{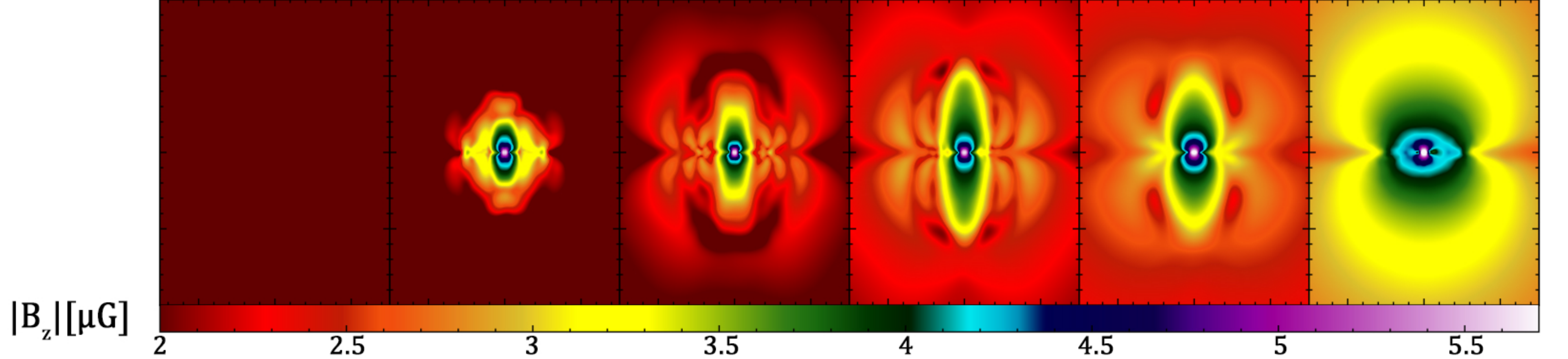}\\
\includegraphics[width=17cm]{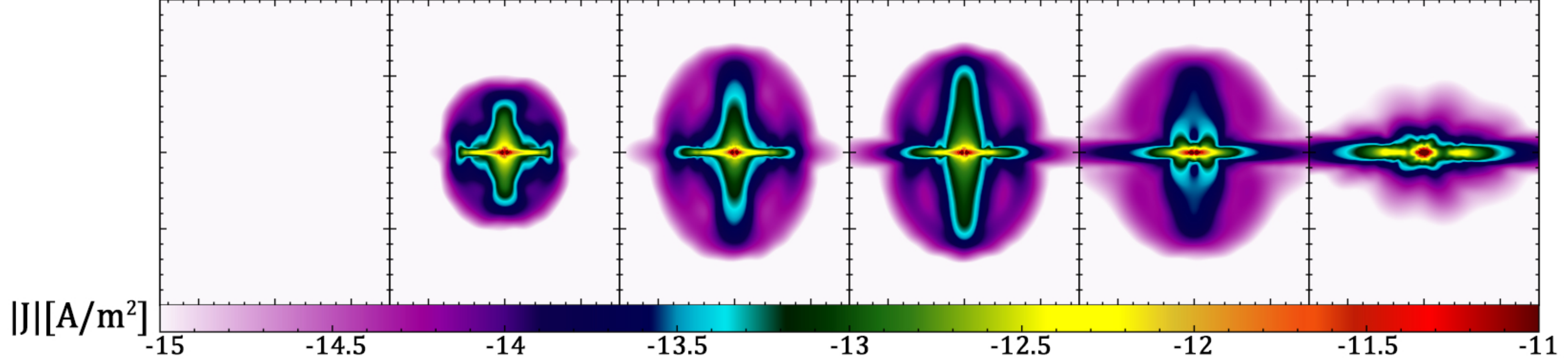}\\
\end{tabular}
    \caption{Result of the magnetized cloud collapse for GDSPH at a resolution of $50^3$ particles with varying magnetic flux ratio $\mu$ going left to right from high to low. We show figures at the time of jet formation (around $t=t_{ff}$), which occurs due to the winding of the magnetic field during the collapse, which produces a magnetic tower structure. The top row shows a rendered face-on slice ($L_{xy}=[2000AU,2000AU]$) of the density $[g/cm^3]$, the rest of the rows show rendered slices through the rotation axis ($L_{xz}=[2000AU,2000AU]$), where the second shows density $[g/cm^3]$, the third show radial velocity $[km/s]$, the fourth show the absolute poloidal magnetic field $[\mu G]$ and the fifth shows the current density $[A/m^2]$, all quantities are shown in logarithmic scale. The pure hydrodynamic run ($\mu=\infty$) of GDSPH becomes gravitationally unstable and is very similar to that of TSPH in \fig~\ref{fig:coll50tsph}. We can see that a jet is launched in the cases of $\mu=75,20,10,5$ while in the case of $\mu=2$ the interchange instability (see \fig~\ref{fig:interchange}) disrupts the disk before jet launching.}
\label{fig:coll50gdsph}
\end{figure*}
\begin{figure*}[]
   \centering
\begin{tabular}{c@{}c}
\includegraphics[width=17cm]{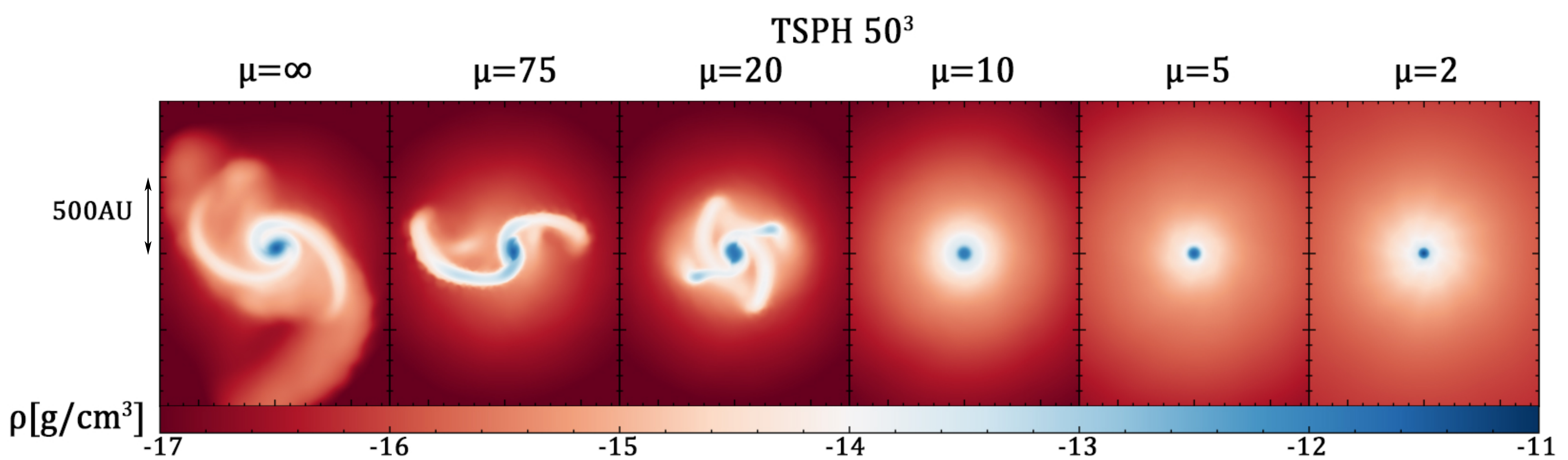}\\
\includegraphics[width=17cm]{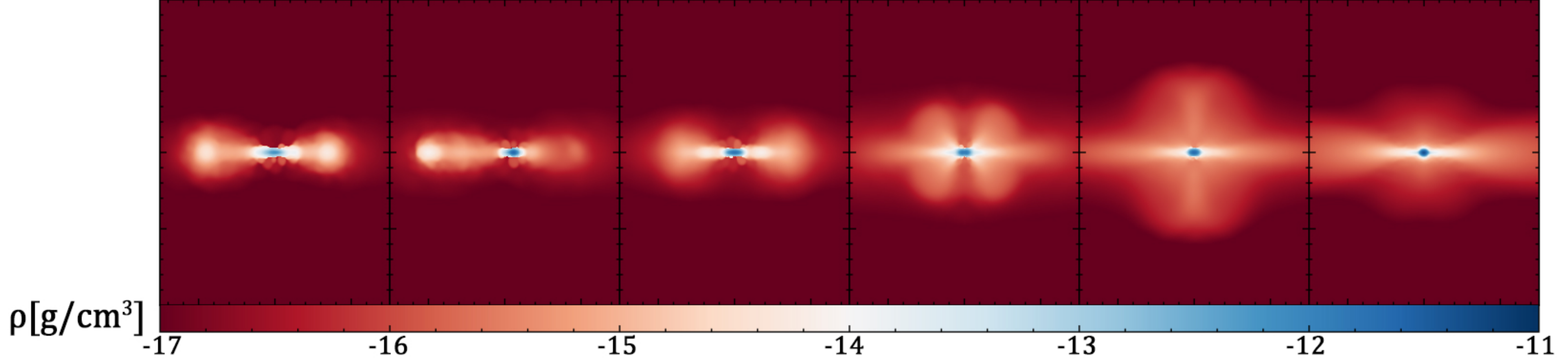}\\
\includegraphics[width=17cm]{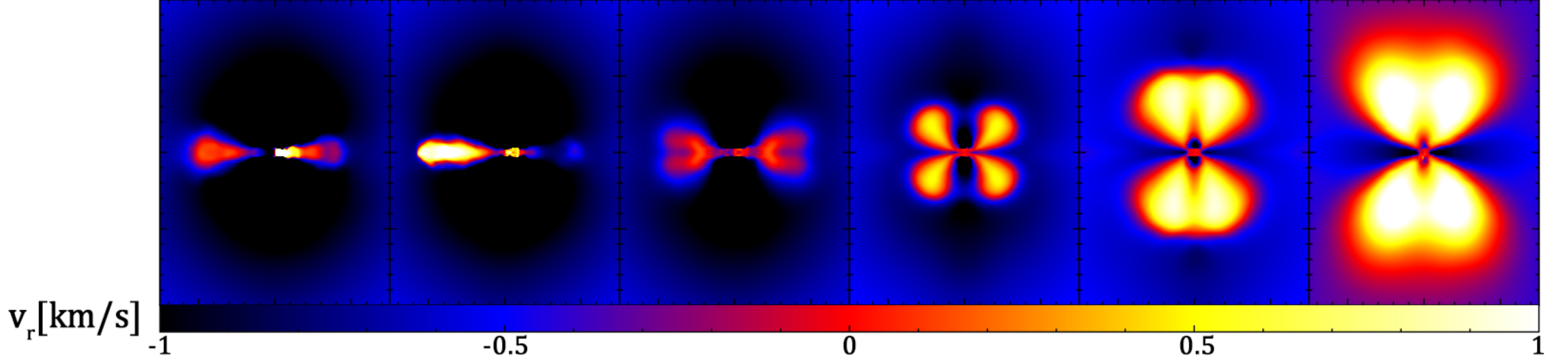}\\
\includegraphics[width=17cm]{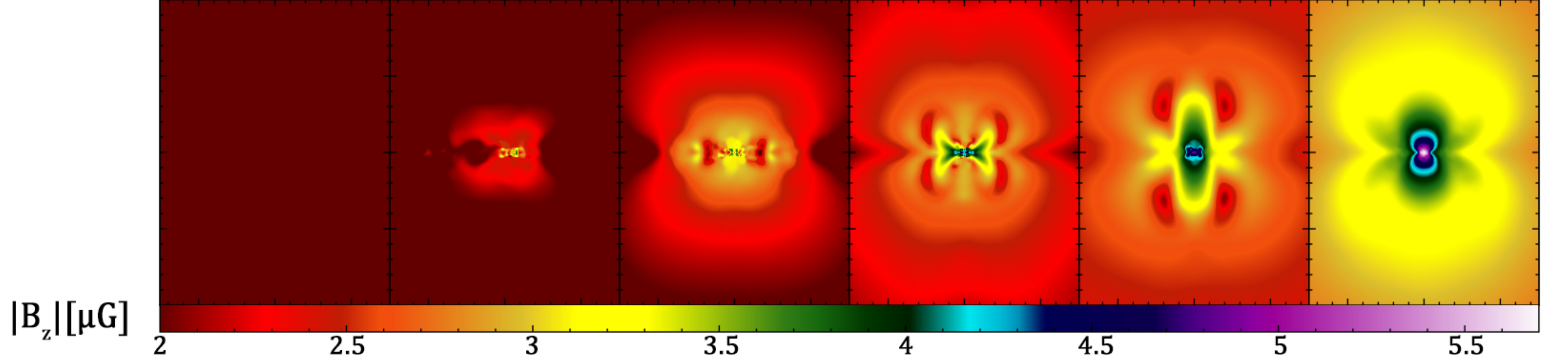}\\
\includegraphics[width=17cm]{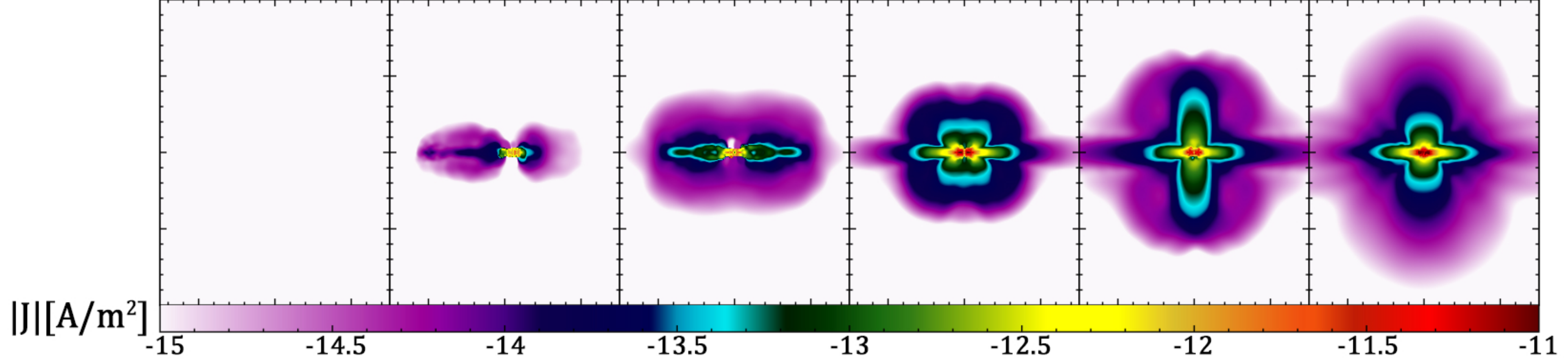}\\
\end{tabular}
    \caption{Result of the magnetized cloud collapse for TSPH at a resolution of $50^3$ particles with varying magnetic flux ratio $\mu$ going left to right from high to low. We show figures at the time of jet formation (around $t=t_{ff}$), which occur due to the winding of the magnetic field during collapse, which produces a magnetic tower structure. The top row shows a rendered face-on slice ($L_{xy}=[2000AU,2000AU]$) of the density $[g/cm^3]$, the rest of the rows show rendered slices through the rotation axis ($L_{xz}=[2000AU,2000AU]$), where the second shows density $[g/cm^3]$, the third show radial velocity $[km/s]$, the fourth show the absolute poloidal magnetic field $[\mu G]$ and the fifth shows the current density $[A/m^2]$, all quantities are shown in logarithmic scale. The pure hydrodynamic run ($\mu=\infty$) of TSPH becomes gravitationally unstable and is very similar to that of GDSPH in \fig~\ref{fig:coll50gdsph}. We can see that TSPH does not form a jet in any of the weak field cases ($\mu=75,20,10$) and there is no sign of a magnetic tower being formed. In the case of $\mu=5$, we can see a jet being launched, where a current dominated magnetic tower has formed, however, the central part of the tower has been completely quenched. The $\mu=2$ case also launches a jet, but collapses faster than in the high resolution case, which effectively leads to easier jet formation.}
    \label{fig:coll50tsph}
\end{figure*}
\begin{figure*}[]
   \centering
\begin{tabular}{c@{}c}
\includegraphics[width=17cm]{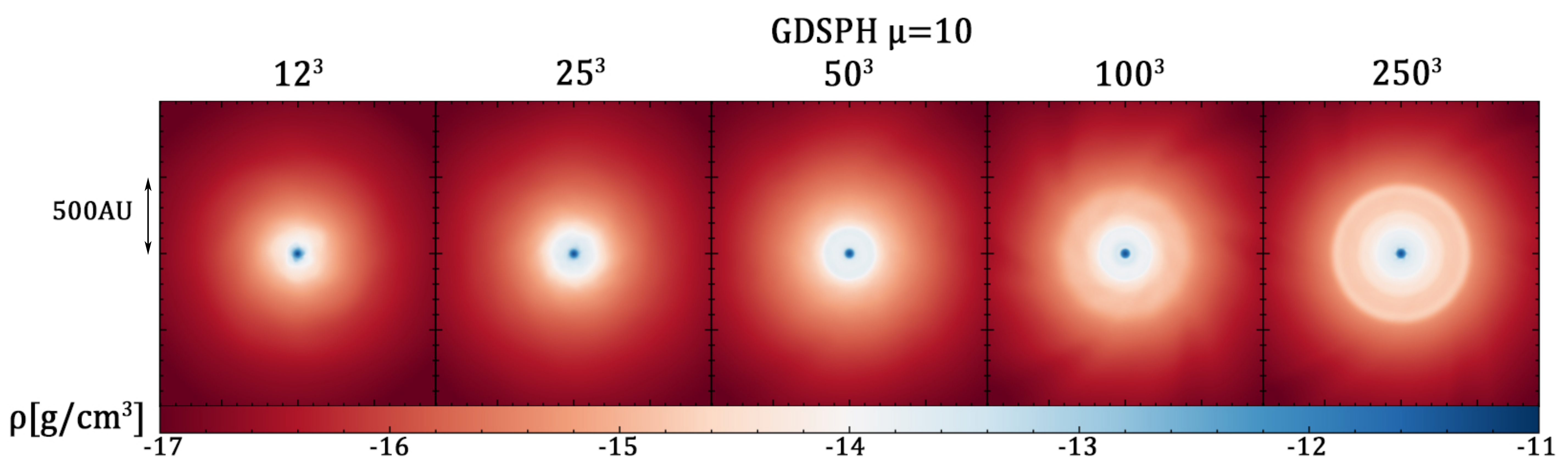}\\
\includegraphics[width=17cm]{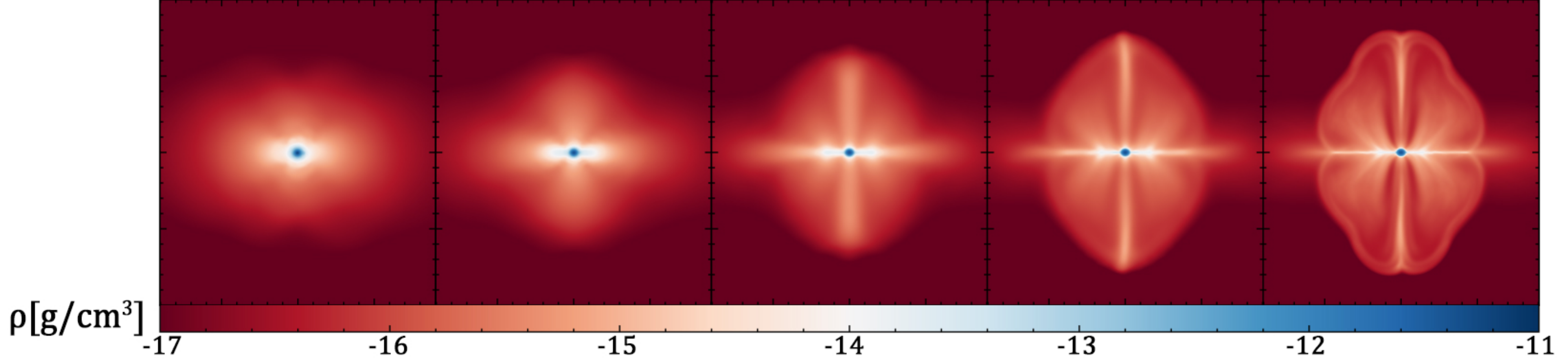}\\
\includegraphics[width=17cm]{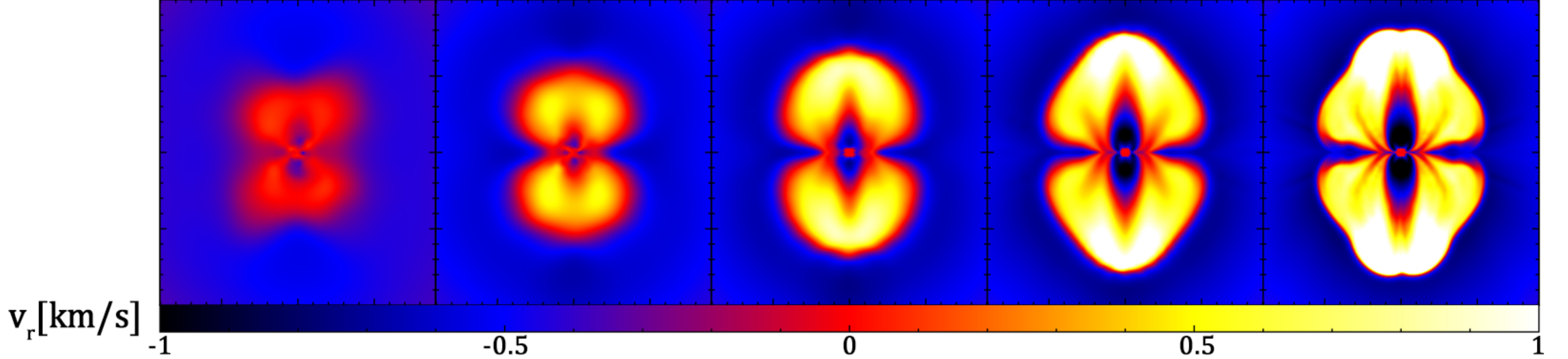}\\
\includegraphics[width=17cm]{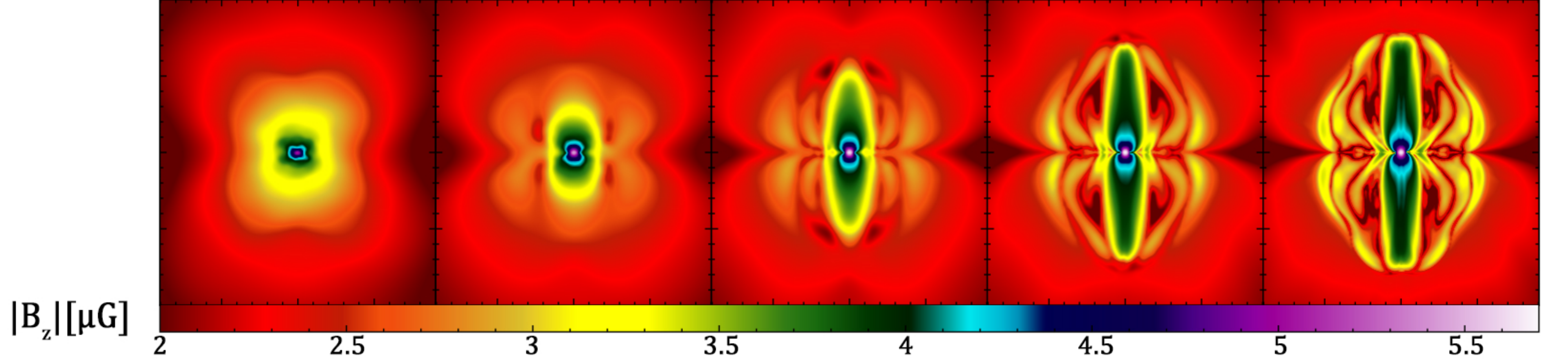}\\
\includegraphics[width=17cm]{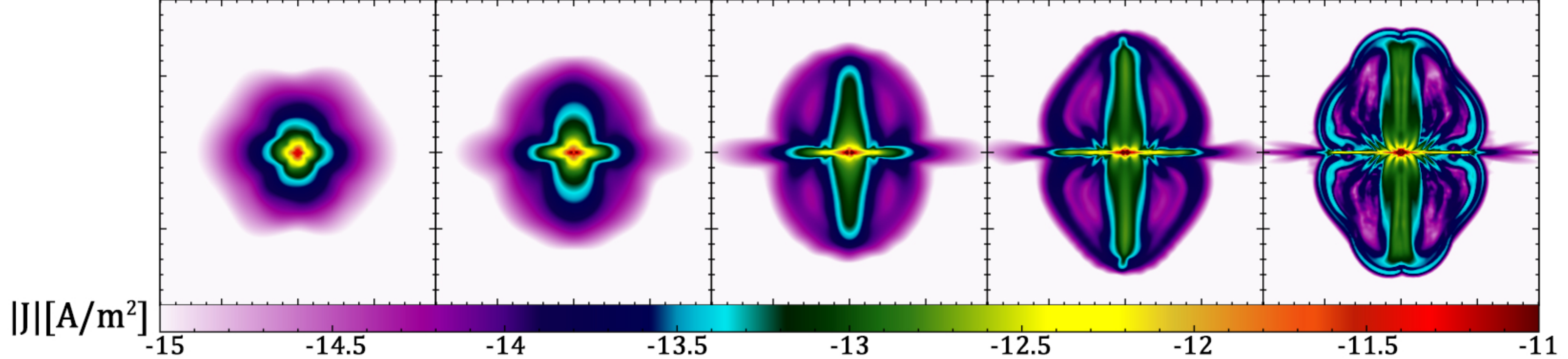}\\
\end{tabular}
    \caption{Result of the resolution study of the magnetized cloud collapse for GDSPH with $\mu=10$. We vary the resolution from left to right, in the initial cloud ($12^3,25^3,50^3,100^3,250^3$) and medium ($10^3,20^3,40^3,80^3,200^3$).We show figures at the time of jet formation (around $t=t_{ff}$), which occur due to the winding of the magnetic field during collapse, which produces a magnetic tower structure. The top row shows a rendered face-on slice ($L_{xy}=[2000AU,2000AU]$) of the density $[g/cm^3]$, the rest of the rows show rendered slices through the rotation axis ($L_{xz}=[2000AU,2000AU]$), where the second shows density $[g/cm^3]$, the third show radial velocity $[km/s]$, the fourth show the absolute poloidal magnetic field $[\mu G]$ and the fifth shows the current density $[A/m^2]$; all quantities are shown in logarithmic scale. Jet formation and a proper magnetic tower can be seen to occur at very low resolution compared to TSPH. The jet structure and magnetic tower further increases in complexity as we increase the resolution. }
      \label{fig:collmu10gdsph}
\end{figure*}
\begin{figure*}[]
   \centering
\begin{tabular}{c@{}c}
\includegraphics[width=17cm]{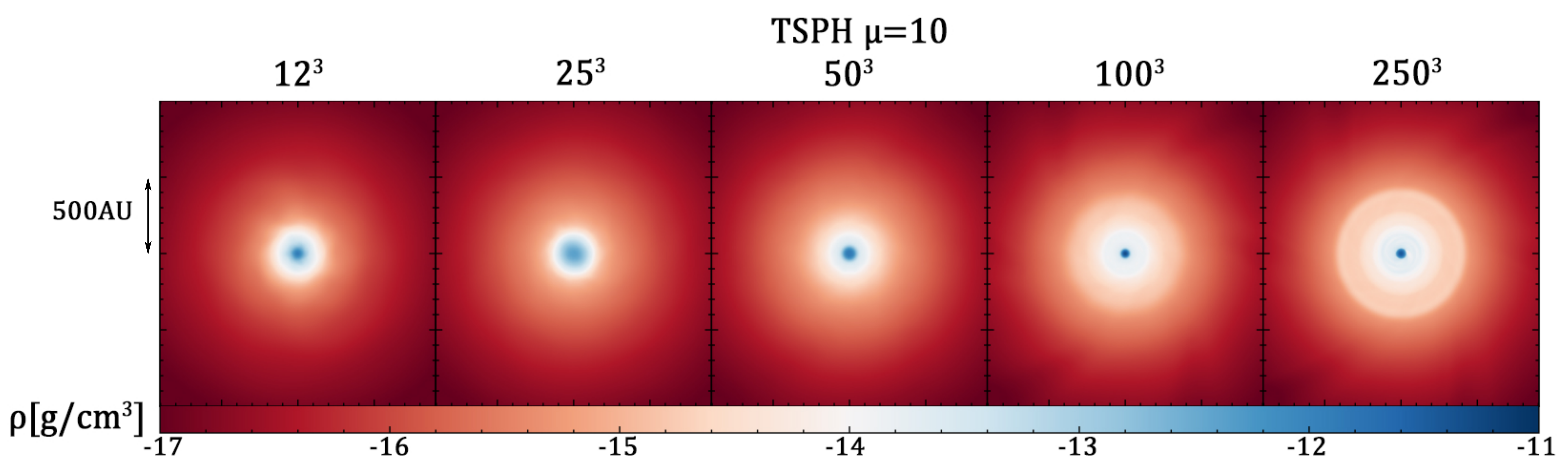}\\
\includegraphics[width=17cm]{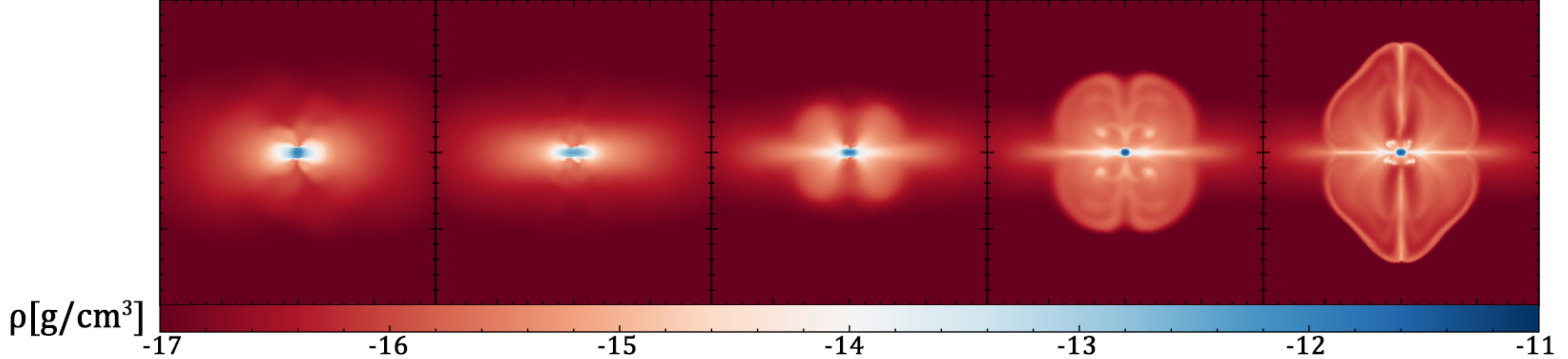}\\
\includegraphics[width=17cm]{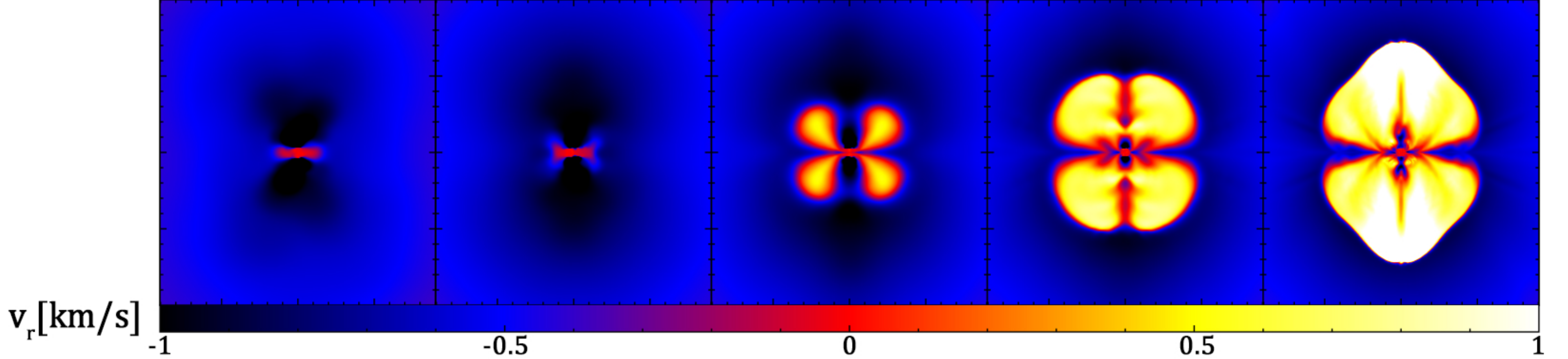}\\
\includegraphics[width=17cm]{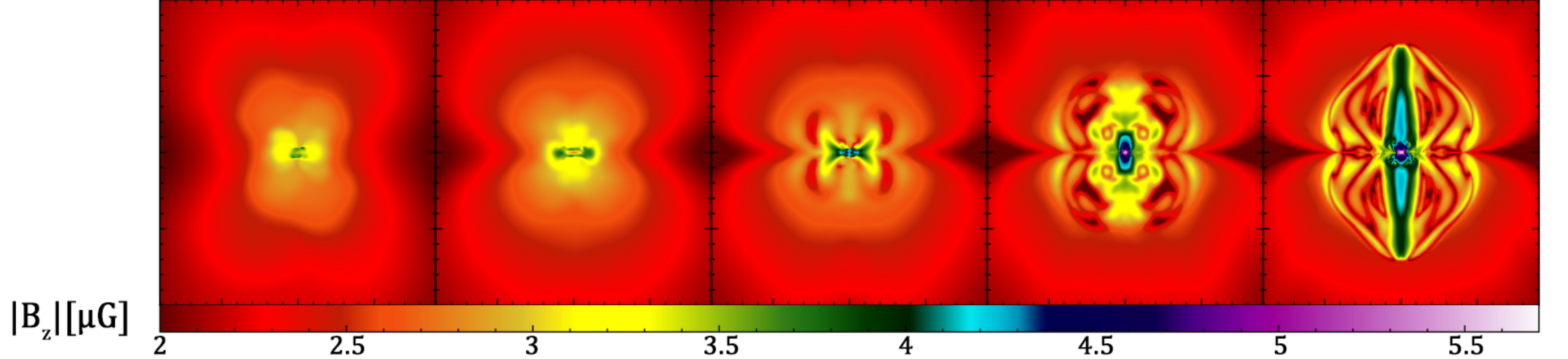}\\
\includegraphics[width=17cm]{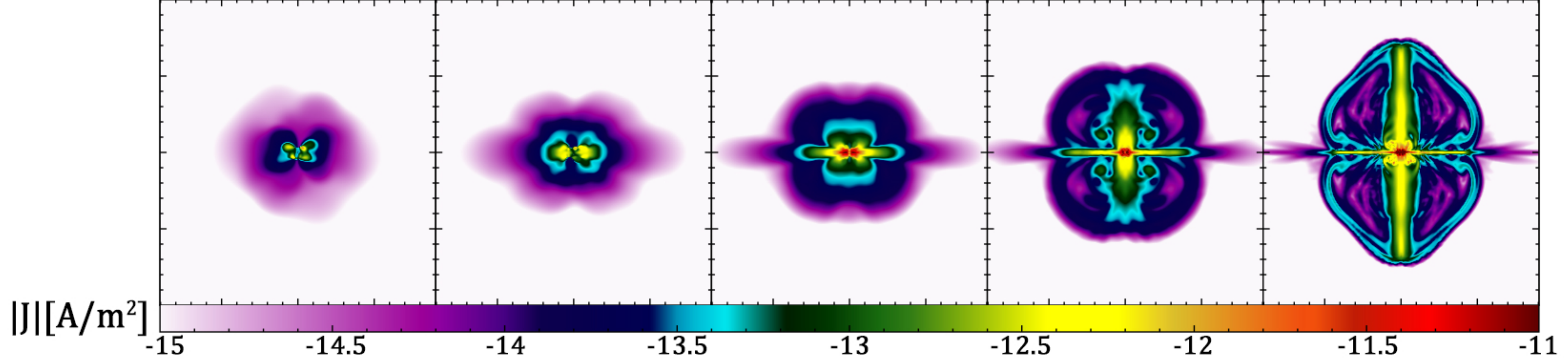}\\
\end{tabular}
    \caption{ Result of the resolution study of the magnetized cloud collapse for TSPH with $\mu=10$. We vary the resolution from left to right, in the initial cloud ($12^3,25^3,50^3,100^3,250^3$) and medium ($10^3,20^3,40^3,80^3,200^3$). We show figures at the time of jet formation (around $t=t_{ff}$), which occur due to the winding of the magnetic field during collapse, which produces a magnetic tower structure. The top row shows a rendered face-on slice ($L_{xy}=[2000AU,2000AU]$) of the density $[g/cm^3]$, the rest of the rows show rendered slices through the rotation axis ($L_{xz}=[2000AU,2000AU]$), where the second shows density $[g/cm^3]$, the third show radial velocity $[km/s]$, the fourth show the absolute poloidal magnetic field $[\mu G]$ and the fifth shows the current density $[A/m^2]$, all quantities are shown in logarithmic scale. We can see that TSPH only forms a collimated jet and a proper magnetic tower at the highest resolution.}
        \label{fig:collmu10tsph}
\end{figure*}
    \begin{figure}
    \centering
\includegraphics[width=\hsize]{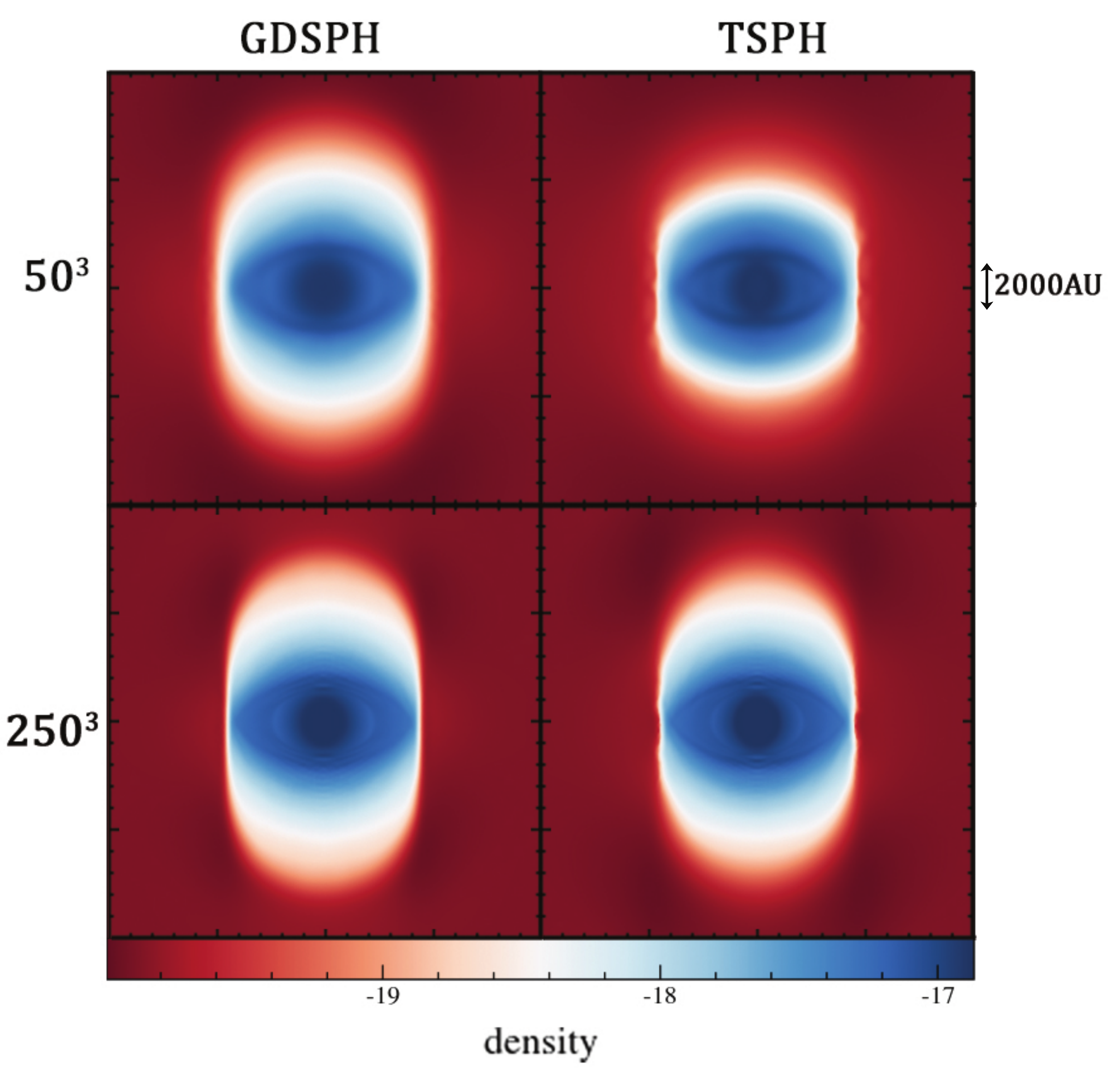}
    \caption{Density rendered slice ($[g/cm^3]$) through the rotation axis, showing the early cloud structure of the strong field case $\mu=2$ before the formation of the disk. As the magnetic field is very strong, accretion will occur primarily along the field lines, creating an elongated cloud structure. Both high resolution cases ($250^3$) of TSPH and GDSPH creates an elongated cloud structure, while in the low-resolution case ($50^3$) only GDSPH forms the same cloud structure. TSPH instead forms a more compact central cloud, which as a consequence collapses faster than the GDSPH case and the high resolution cases.}
    \label{fig:cloudstruct}
\end{figure}
    We followed the setup outlined in \cite{2008A&A...477....9H} and \cite{2016MNRAS.455...51H} and set up a 3D periodic box $L = [0.15 \rm pc,0.15pc,0.15pc]$. A cloud was initiated with a radius of $R_c=0.015\ \rm pc$ and a mass of 1 solar mass ($M_c=1 \rm M_{\odot}$), within a surrounding medium that has 360 times lower density than the cloud ($\rho_{out}=M_c/(360 V_c)$). The cloud was put in rotation with an orbital time of $P=4.7\times 10^5 \ \rm yr$, which corresponds to a kinetic over potential energy ratio of about $E_K/E_P\approx0.045$. This is a higher ratio compared to the peak value of $0.02$ from the observed distribution of rotation rates in molecular clouds ($E_K/E_P \in (0.002,1.4)$) \citep{1993ApJ...406..528G}. A constant magnetic field $B_0$ was initialized in the direction of the angular momentum vector ($\hat{\textbf{z}}$), and we varied the strength in accordance to different mass-to-flux ratios. The mass-to-flux ratio $\mu$ is relative to the critical mass-to-flux ratio, $(M_{c}/\Phi)_{crit}$, in which the cloud is fully supported by magnetic forces against gravity, that is,
\begin{equation}
\mu=\left(\frac{M_c}{\Phi}\right)/\left(\frac{M_c}{\Phi}\right)_{crit} ,
\end{equation}
\begin{equation}
    \left(\frac{M_c}{\Phi}\right)=\frac{M_c}{\pi R_c^2 B_0}, \quad \left(\frac{M_c}{\Phi}\right)_{crit}=\frac{c_1}{3\pi}\sqrt{\frac{5}{G}} .
\end{equation}
Here, $c_1=0.53$ is a parameter that can be determined numerically \citep{1976ApJ...210..326M}. We then get the corresponding initial magnetic field:
\begin{equation}
B_0=\frac{610}{\mu} \ [\mathrm{ \mu G }] .
\end{equation}
The thermal pressure is determined by the following barotropic EOS, 
\begin{equation}
    P=c_{s,0}^2\rho\sqrt{1+(\rho/\rho_0)^{4/3}},
\end{equation}
with $\rho_0=10^{-14} \  \mathrm{g \ cm^{-3}}$ and $c_{s,0}=0.2 \ \mathrm{km \ s^{-1}}$. We looked at six different magnetic flux ratio values in our simulations, from weak to high ($\mu=\infty,75,20,10,5,2$). These were run with a moderate resolution of $50^3$ in the cloud, which corresponds to about $40^3$ particles in the low density medium, same as in the setup of \cite{2016MNRAS.455...51H}. These six cases were run with both GDSPH and TSPH until the core has fully collapsed, close after the time of jet launching, which typically occurs when the maximum density hits a value in between $\rho = 10^{-12}\leftrightarrow10^{-11} g/cm^3$. This occurs near the free fall time $t_{ff}=\sqrt{\frac{3}{2\pi G \rho}}\approx 4 \times 10^4 \ \rm yr$, at around $t=1.05t_{ff}\leftrightarrow1.3 t_{ff}$ depending on resolution/initial magnetic field strength. No sink particles were used in any of our simulations. The results of these simulations can be seen in \fig~\ref{fig:coll50gdsph} (GDSPH) and \fig~\ref{fig:coll50tsph} (TSPH). 
\\ \\
\begin{figure}
    \centering
\begin{tabular}{c@{}c}
\includegraphics[width=\hsize]{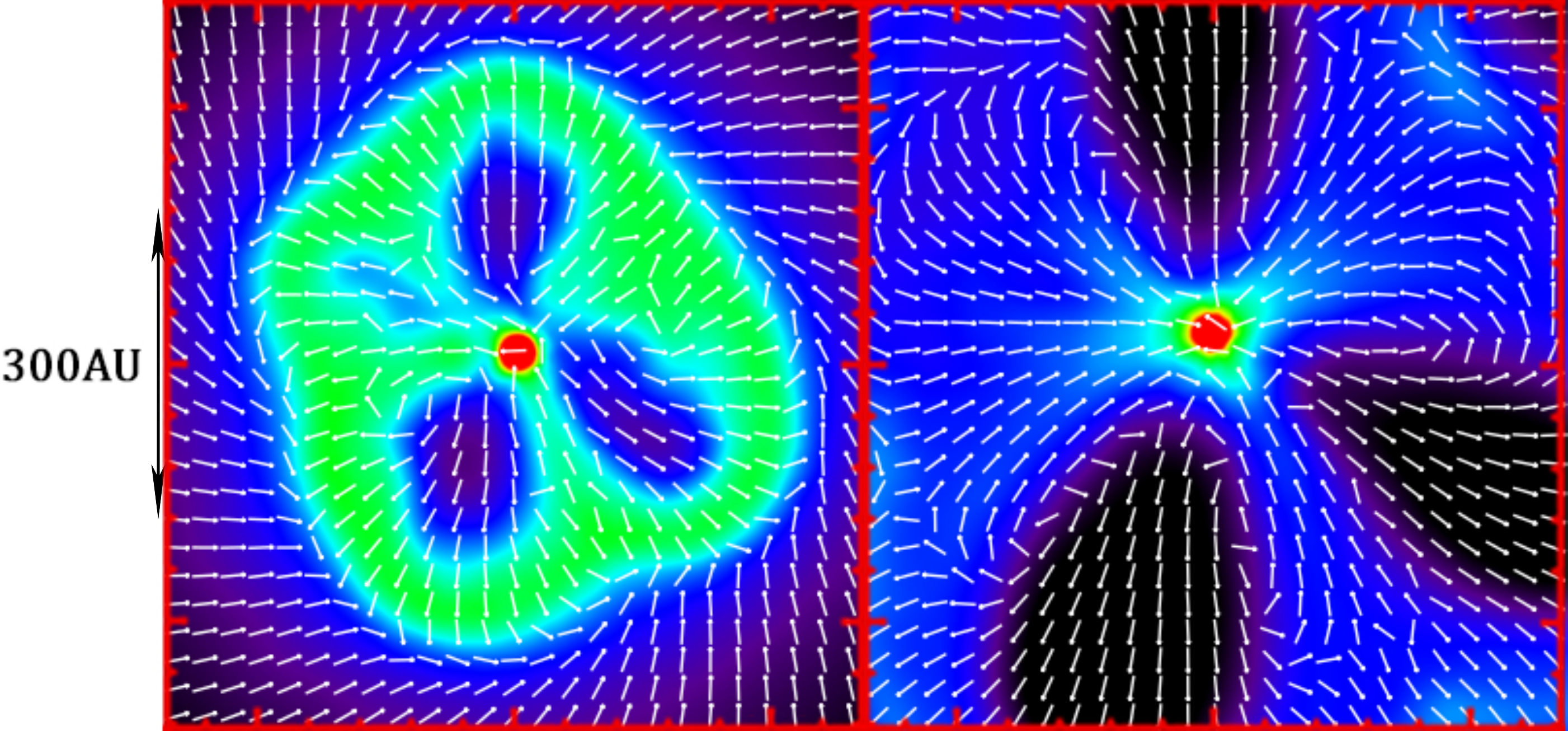}
\end{tabular}
    \caption{Magnetic interchange instability in the strong field case ($\mu=2$) for the GDSPH simulation with $50^3$ resolution. The left panel shows a zoom-in of the disc structure seen face-on in \fig~\ref{fig:coll50gdsph} and the right panel depicts the same region at a later time. The white arrows show the direction of velocity and the colour scale indicates density. The instability launches magnetized bubbles in the azimuthal direction. At later times we can see that the central star starts to accrete again along the filamentary structure. These figures can be compared to the results from \citet{2012ApJ...757...77K}}
    \label{fig:interchange}
\end{figure}
The pure hydrodynamic runs ($\mu=\infty$) of both GDSPH and TSPH become gravitationally unstable and the resulting evolution is very similar (see \fig~\ref{fig:coll50gdsph} and \fig~\ref{fig:coll50tsph}). For GDSPH, we see in \fig~\ref{fig:coll50gdsph} that jet launching can be seen in the weak field regime ($\mu= 75,20,10$), although it is very short-lived in the case of $\mu=75$. It is clear from the poloidal magnetic field and the current density (the fourth and fifth row in \fig~\ref{fig:coll50gdsph}) that we have a developed magnetic tower in all these three cases. This is a remarkable achievement, especially for SPMHD, since the amplification of the magnetic field can easily be quenched by numerical dissipation. The jet strength, morphology and velocity structure resemble those in \cite{2016MNRAS.455...51H} with the same resolution using the MFM/MFV method with more complex gradient operators and Riemann solvers. In contrast, for TSPH (\fig~\ref{fig:collmu10tsph}) we see neither jet launching nor a magnetic tower in the weak-field regime with the fiducial resolution of $50^{3}$. This is largely due to numerical dissipation which suppresses field amplification and hinders the formation of a jet.  At the time of jet launching, fragmentation of the disc occurs in the two weakest cases $\mu=75$ and $\mu=20$ for TSPH, as the magnetic field is too weak to support the disc. For GDSPH, it does not occur until a later time in the simulation, however, the exact time of fragmentation is heavily dependent on other dissipation terms such as artificial viscosity.
\\ \\
\begin{figure*}
    \centering
\includegraphics[width=\hsize]{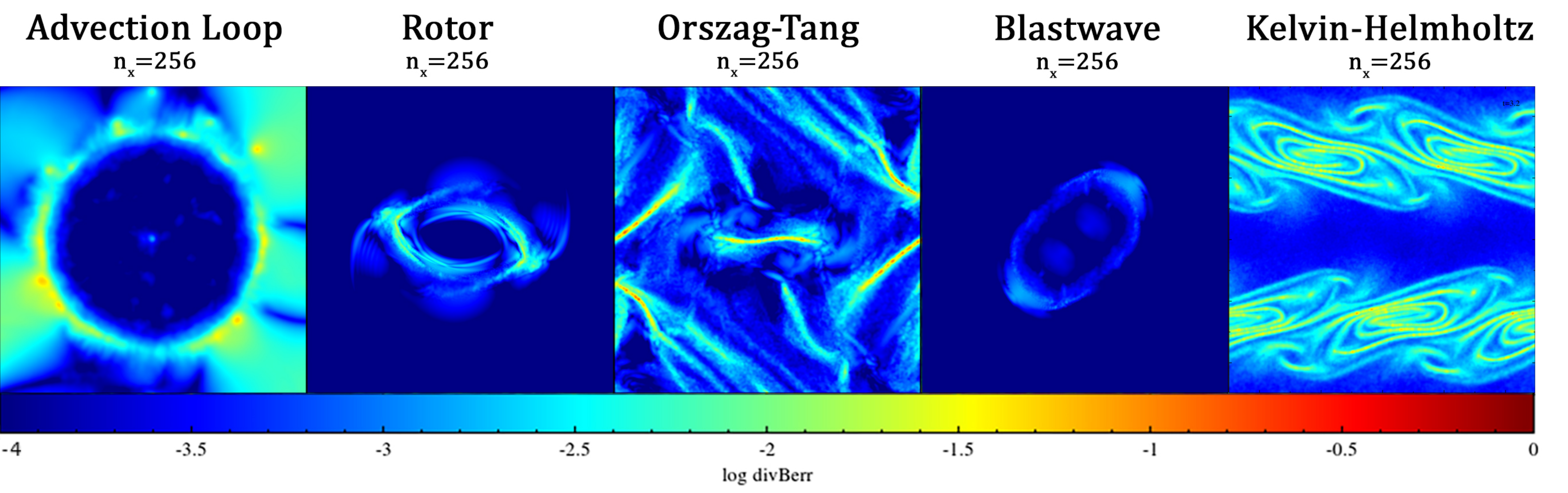}
    \caption{Normalized divergence error ($\epsilon_{divB} \equiv h|\nabla\cdot \Bv|/|\Bv|$, \eq~\ref{eq:divBerr}) of the different cases (all in 3D) with GDSPH. From left to right: First: advection of a current loop with a density ratio of $\Delta=2$ at $t=20$ (Section~\ref{subsec:advloop}), Second: MHD rotor ($n_x=256$) at $t=0.15$ (Section~\ref{subsec:rotor}). Third: Orszag-Tang vortex ($n_x=256$) at $t=0.5$ (Section~\ref{subsec:orz}). The MHD blastwave ($n=256^3$) at $t=0.02$ (Section~\ref{subsec:mhdblast}). The MHD Kelvin-Helmholtz instability ($n_x=256$) with an initial sharp contact discontinuity at $t=3.2$ (Section~\ref{subsec:kh}).}
    \label{fig:divBerr}
\end{figure*}
For the more magnetized case with $\mu=5$ we can see that TSPH successfully develops a jet, however, closer inspection on $|B_{z}|$ in \fig~\ref{fig:coll50tsph} shows that the central portion of the magnetic tower is much less developed, with an order of magnitude smaller strength than the same run with GDSPH. Jet launching is also seen in the GDSPH case with $\mu=5$ with the magnetic tower intact, albeit weaker and less collimated than in the $\mu=10$ case. For $\mu=2$, the collapse proceeds differently in GDSPH and TSPH from an early stage. This can be seen in \fig~\ref{fig:cloudstruct}, which shows the density structure of the collapsing cloud in an early stage with different resolutions. As the magnetic field is very strong, accretion will occur primarily along the field lines, creating an elongated cloud structure. While at high resolution ($250^3$) both GDSPH and TSPH runs converge to the correct structure, at the fiducial resolution of $50^{3}$, only GDSPH shows an elongated cloud. The cloud collapses faster in TSPH, likely due to excessive dissipation.
\\ \\ 
For $\mu=2$, we can see that GDSPH does not produce a coherent jet (\fig~\ref{fig:coll50gdsph}). This is mainly due to the disk being disrupted by the magnetic interchange instability. This instability occurs due to an accumulation of magnetic flux near the accreting protostar, where the magnetic flux that would have been dragged into the protostellar core is redistributed to the surrounding medium by dissipative effects (AR). This builds a large magnetic pressure gradient which, together with the twisted magnetic field, eventually launches highly magnetized bubbles in the azimuthal direction. A density rendering together with a velocity map of the $\mu=2$ case after the launch of the magnetized bubbles is shown in \fig~\ref{fig:interchange}. This is similar to the results seen in simulations using the AMR code {\sc Enzo} \citep{2012ApJ...757...77K} and in SPH simulations with {\sc Phantom} \citep{2017arXiv170607721W}. We would like to stress that, unlike the SPH runs in \cite{2016MNRAS.455...51H}, the disk disruption is not due to divergence errors, but instead a consequence of the magnetic dissipation. At later times (for example, right panel of \fig~\ref{fig:interchange}), we can see that the protostellar core remains centrally located, which indicates good divergence control and angular momentum conservation. As the formation of the interchange instability is driven by the redistribution of magnetic flux, it can depend heavily on the choice of AR prescription and the use of sink particles. \cite{2017arXiv170607721W} observed similar magnetic bubbles with the same AR prescription as ours, while other tested AR prescriptions did not launch magnetic bubbles. However, all other works that produce interchange instabilities use sink particles, which can artificially redistribute the flux as matter is accreted by the sink, while leaving the magnetic field close to the sink intact. The development of the interchange instability in our simulations without sink particles might indicate a more physical origin of the effect. Additional work will need to be done to determine if this is in fact a real effect or a consequence of the numerical scheme.
\\ \\
To investigate the convergence with resolution, we simulated the $\mu=10$ case across different resolutions ($12^3$,$25^3$,$50^3$,$100^3$ and $250^3$) for both methods. The results are shown in \fig~\ref{fig:collmu10gdsph} for GDSPH and \fig~\ref{fig:collmu10tsph} for TSPH. In GDSPH, resolved jet structures and fully developed magnetic towers are already evident in cases with $25^3$ resolution, and which increase in complexity as we increase the resolution. A weak outflow appears even in the lowest resolution of $12^3$. In contrast, the runs with TSPH shows slow convergence. The structure of the magnetic field is severely distorted, and magnetic tower and proper collimated jet are not developed in all cases except the highest resolution. Again, we note that our GDSPH results are very comparable to the MFM/MFV runs in \cite{2016MNRAS.455...51H}, both in terms of jet properties and converging speed. The magnetic tower structure is also qualitatively similar to the cloud collapse simulation in the weak field regime from the moving-mesh code {\sc Arepo}  \citep{2011MNRAS.418.1392P}, although with a slightly different initial setup. We should note that the collapse of the $12^3$ is artificially suppressed and contract much slower then what is expected. This is because the local Jeans mass is not fully resolvable in these simulations. \cite{1997MNRAS.288.1060B} estimated that around $3\cdot10^4$ particles where required to correctly resolve the Jeans mass in similar collapse cases\footnote{The number of particles required to resolve the Jeans mass is presumably even higher in our simulation as \cite{1997MNRAS.288.1060B} used a smaller neighbour number ($N_{neigh}=50$) and neglected magnetic fields.}. The effect can partly be seen in the $25^3$ case as well, especially at later times. However, in this case, the cloud collapsing has a similar evolution up to the time of jet-launching as the higher resolution cases.

\section{Discussion}
\label{sec:discussion}

\begin{figure}
    \centering
\includegraphics[width=\hsize]{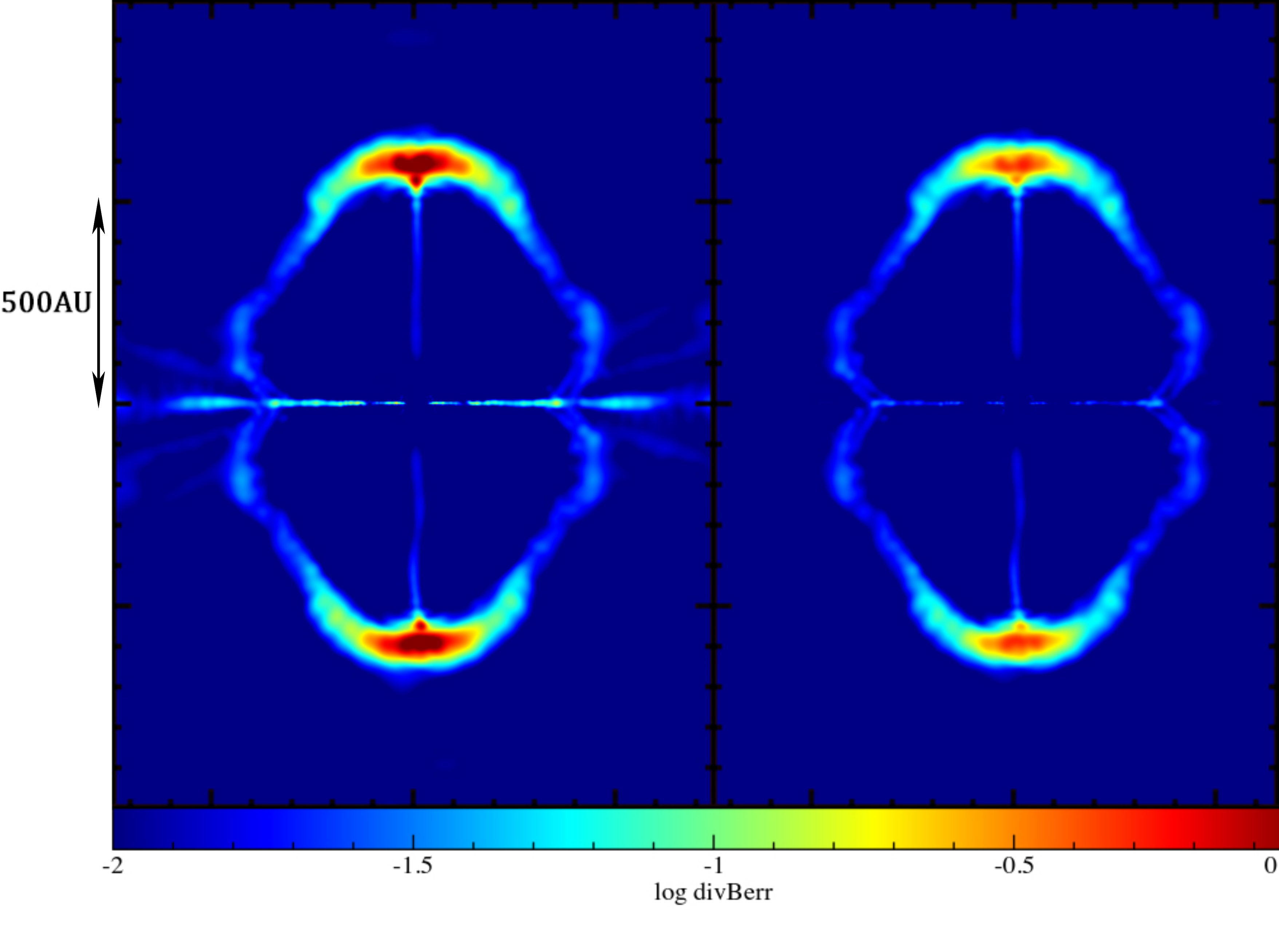}
    \caption{Normalized divergence error of the magnetized cloud collapse simulation with GDSPH at the jet launching time, $t\approx t_{ff}$. The mass-to-flux ratio is $\mu=10$ and the resolution is $100^3$. {\it Left:} normalized divergence error as in (\eq~\ref{eq:divBerr}), $\epsilon_{divB} \equiv h|\nabla\cdot \Bv|/|\Bv|$; {\it Right:}  divergence error normalized to total gas pressure, $\epsilon_{divB}/\sqrt{(1+\beta)}=h|\nabla\cdot \Bv|/\sqrt{({\Bv}^2 + 2P)}$, where $\beta=2P/\Bv^{2}$ is the plasma beta. }
    \label{fig:divBerrcollapse}
\end{figure}

In this paper, we present an SPMHD method, which utilizes the Geometric Density average force discretization (GDSPH) of the MHD equations. GDSPH has been shown in previous work to greatly improve the accuracy near density discontinuities and eliminate the surface tension problem. We show that MHD also benefits from this method. For a large part, the standard test problems (Section 3.1-3.6) both GDSPH and TSPH handle the problems very well, and the differences between the two methods are minimal. However, when the problem involves mixing such as in the case of Kevin-Helmholtz instabilities, GDSPH shows clear advantages. This is somewhat expected, and in agreement with earlier studies without magnetic fields \citep{2017MNRAS.471.2357W}. However, when we apply the method in the astrophysical test of a collapsing magnetized cloud, we see that GDSPH leads to a significant improvement. GDSPH not only realistically captures the development of a magnetic tower and jet launching in the weak field regime ($\mu \gtrsim 10$), but also exhibits a fast improvement in the complexity and structure of the jet with increased resolution. In contrast, TSPH only manages to launch jets in the strong field regime with a resolution of $50^3$ and only develop a collimated jet in the highest resolution runs of the $\mu=10$ case. We also show that, in the strong field regime, GDSPH converges better than TSPH in accretion time and in the outer cloud structure. The results of TSPH is in agreement with previous studies using TSPH \citep{2016MNRAS.455...51H,2011MNRAS.417L..61B} which used the same IC as this work, and other studies \citep{2007MNRAS.377...77P,2012MNRAS.423L..45P} which employ a smaller initial cloud rotation ($E_K/E_P = 0.005$).
\\ \\ 
Overall, our new method shows improved or comparable results to previous SPMHD implementations such as in {\sc Phantom} \citep{2018PASA...35...31P} and in {\sc Gizmo} \citep{2016MNRAS.455...51H}. In many test cases and particularly in the cloud collapse case, GDSPH produces qualitatively very similar result to that of the MFM/MFV method and achieves similar convergence speed. It is worth noting that all the simulations are run in 3D with the code default parameters listed in Section~\ref{sec:tests} without any adjustment by hand to specific problems. The success of GDSPH can most likely be attributed to the reduction of SPH ``$E_{0}$ errors'' (\eq~\ref{eq:e0error}) and linear errors by the geometric density average force formulation. As discussed in Section~\ref{subsec:sphdisc}, the advantage of such discretization is more evident when the density gradient is large, such as in the cloud collapse case.
\\ \\
Using the constrained hyperbolic divergence cleaning scheme with variable cleaning speed from \citet{2016JCoPh.322..326T}, we can keep the divergence error low in all cases. The mean normalized divergence error, $\langle\epsilon_{divB}\rangle = \langle h|\nabla\cdot \Bv|/|B|\rangle$, is typically of order $10^{-5} - 10^{-3}$. In \fig~\ref{fig:divBerr}, we show the normalized divergence error maps for several test problems. Again we see that the divergence cleaning works extremely well here, the maximum error is generally around $10^{-2}$. Comparing to \citet{2016MNRAS.462..576H} (their Figure 4), we find that the errors are smaller than their MFM simulations with the \citet{2002JCoPh.175..645D} cleaning in general, with an exception for the outskirt of the advection loop (where the magnetic field is essentially zero and thus not important for the result). The improvement is probably due to the more advanced constrained cleaning method \citep{2016JCoPh.322..326T}. The normalized divergence error for the $\mu=10$ cloud-collapse case at the jet launching time is shown in \fig~\ref{fig:divBerrcollapse}. Here, the divergence cleaning still performs very well in the disc, along the jets and for the majority of the regions where the outflow interacts with the ambient gas, especially when the divergence error is compared to the total gas pressure (right panel). The result is similar to the Dedner cleaning in \citet{2016MNRAS.462..576H}, although our error is somewhat larger at the tip of the jets where the gas is shocked. However, we note that the comparison is not direct in this case as the jets may develop differently.  Overall, the result from cleaning is still worse than the constrained transport or constrained gradient schemes \citep{2016MNRAS.462..576H}. For SPMHD, as shown in \cite{2012JCoPh.231.7214T}, divergence errors can be reduced to machine precision (or more practically to a certain tolerance value) using cleaning, with the help of a sub-cycling routine. However, local adjustments are required to determine the number of iterations for each particle to efficiently subcycle the cleaning in the simulation. This is because certain regions are more affected than others and because divergence is spread to nearby neighbours. Conceivably, if vector potentials \citep{2015JCoPh.282..148S} could work for a wider range of problems this could be an interesting avenue as well. However, the exploration of these methods in detail is beyond the scope of this work.
\\ \\
From the tests, we can see that using a lower artificial resistivity coefficient ($\alpha_B=0.5$) than ($\alpha_B=1$) \citep{2018PASA...35...31P} works well for all cases. However, the choice of artificial resistivity switch still remains somewhat ad-hoc as it is difficult to accurately detect the MHD discontinuity. A Godunov SPH scheme \citep{2011MNRAS.418.1668I} solves this problem by replacing artificial resistivity with Riemann solvers, which have been shown to produce minimal artificial diffusivity. However, this brings with it an increase in computational cost and it is unclear if the extra cost is worth it. Further improvements in the convergence of SPMHD may be found in the use of integral-based gradient estimates, which have been shown to be more accurate and less noisy than the standard SPH gradient estimate \citep{2012A&A...538A...9G,2015MNRAS.448.3628R,2017A&A...606A..78C}. This could be especially beneficial for modeling subsonic turbulent flows \citep{2016ApJ...831..103V}. This gradient estimate can easily be implemented within the GDSPH framework and will be investigated in future work. 
\\ \\
Meshless methods (SPH and MFM) were recently explored in local 3D simulations of the magnetorotational instability (MRI) \citep{2019arXiv190105190D}. The authors found that in the vertically stratified MRI simulations, SPH developed an unphysical state with strong toroidal field components and no sustained turbulence. In a forthcoming paper (Wissing et al. {\it in prep.}) we will show that GDSPH do not show this unphysical behavior and that they reproduce the characteristic periodic azimuthal magnetic field pattern (butterfly diagram) of the stratified MRI. 
\\ \\

\section*{Acknowledgements}
We thank the anonymous referee for the useful comments and suggestions which have improved the quality and clarity of the paper. We thank James Wadsley, Tom Quinn, Max Grönke and Ben Keller for very helpful and interesting discussions. The simulations were performed using the resources from the National Infrastructure for High Performance Computing and Data Storage in Norway, UNINETT Sigma2, allocated to Project NN9477K. We also acknowledge the support from the Research Council of Norway through NFR Young Research Talents Grant 276043.  

\bibliographystyle{aa}
\bibliography{references}

\end{document}